\documentstyle[12pt,aaspp4]{article}

\newcommand \myfig[1] {\includegraphics{#1} \vspace{10cm}}
 
\received{1995 March 7}

\begin{document}

\title{Non-thermal radiation of cosmological $\gamma$-ray bursters}

\author{M.V. Smolsky and V.V. Usov} 

\affil{Department of Condensed-Matter Physics,
Weizmann Institute of Science, Rehovot
76100, Israel} 

\authoremail{fnusov@weizmann.weizmann.ac.il}

\date{}

\begin{abstract}

We use one-and-a-half
dimensional particle-in-cell plasma simulations
to study the interaction of a relativistic, strongly magnetized wind 
with an ambient medium. Such an interaction is a plausible
mechanism which leads to generation of cosmological $\gamma$-ray
bursts. We confirm the idea of M\'esz\'aros and Rees
(1992) that an essential part (about 20\%) of the energy that is lost by
the wind in the process of its deceleration may be transferred to 
high-energy electrons and then to high-frequency (X-ray and 
$\gamma$-ray) emission. We show that in the wind frame the spectrum 
of electrons which are accelerated at the wind front and move ahead
of the front is nearly a two-dimensional relativistic
Maxwellian with a relativistic temperature 
$T=m_ec^2\Gamma_{_T}/k\simeq 6\times 10^9\Gamma_{_T}$~K,
where $\Gamma_{_T}$ is equal to $200\Gamma_0$
with the accuracy of $\sim 20$\%, and $\Gamma_0$ is the Lorentz 
factor of the wind, $\Gamma_0\gtrsim 10^2$ for winds outflowing from
cosmological $\gamma$-ray bursters. Our simulations point to
an existence of a high-energy tail of accelerated electrons with
a Lorentz factor of more than $\sim 700 \Gamma_0$.
Large-amplitude electromagnetic waves are generated by 
the oscillating currents at the wind front. The mean field of
these waves ahead of the wind front is an order of magnitude
less than the magnetic field of the wind. High-energy electrons 
which are accelerated at the wind front and injected
into the region ahead of the front 
generate synchro-Compton radiation in the fields of large-amplitude 
electromagnetic waves. This radiation closely resembles
synchrotron radiation and can reproduce  the
non-thermal radiation of $\gamma$-ray bursts observed in the Ginga 
and BATSE ranges (from a few keV to a few MeV). Synchrotron
photons which are generated in the vicinity of the wind front may
be responsible for the radiation of $\gamma $-ray bursts in the EGRET 
energy range above a few ten MeV. The spectrum of 
$\gamma$-ray bursts in high-energy $\gamma$-rays may extend,
in principle, up to the maximum energy of the accelerated electrons 
which is about $10^{13}(\Gamma_0/10^2)^2$~eV in the frame of
the $\gamma$-ray burster.

\keywords{acceleration of particles --- radiation mechanism:
nonthermal --- gamma-ray: bursts --- gamma-rays: theory}

\end{abstract}

\newpage
 
\section{Introduction}
\label{Introduction} 
Many ideas about the nature of $\gamma$-ray bursts have
been discussed during last 25 years after their discovery (for
a review, see \cite{Blaes94}; \cite{Harding94}; \cite{Hartman95}; 
\cite{Dermer95}; \cite{FM95}; \cite{Greiner98}; \cite{Piran98}).
Among these ideas, there was a suggestion that 
the sources of $\gamma$-ray bursts (GRBs) are at cosmological 
distances, i.e. at a redshift $z\sim 1$ (\cite{Usov75}; 
\cite{vdBerg83}; \cite{Paczynski86}; \cite{Goodman86};
\cite{Eichler89}). After the BATSE data became available 
(\cite{Meegan92}, 1994), 
the idea of a cosmological origin of GRB sources has come 
to be taken very seriously (e.g., \cite{Paczynski91}; \cite{FM95}).
Recent detections of absorption and emission features at a redshift
$z=0.835$ in the optical afterglow of GRB 970508 (\cite
{Metzger97}) and at redshift $z=3.42$ in the host galaxy of GRB 
971214 (\cite{Kulkarni98}) clearly demonstrate that at least
some of the GRB sources lie at cosmological distances.
A common feature of all
acceptable models of cosmological $\gamma$-ray bursters is that a
relativistic wind is a source of GRB radiation. The Lorentz factor,
$\Gamma_0$, of such a wind is about $10^2-10^3$ or
even more (e.g., \cite{Fenimore93}; \cite{Harding97}).
A very strong magnetic field may be in the plasma 
outflowing from cosmological $\gamma$-ray bursters (\cite{Usov92}, 
1994a,b; \cite {Thompson93}; \cite{Blackman96}; \cite{Vietri96};
\cite{Katz97}; \cite{Meszaros97}; \cite{Dai98}). It was pointed
out (\cite{Meszaros92}, 1993; \cite{Rees92}) that the kinetic energy
of relativistic winds may be converted into non-thermal
radiation of GRBs when these winds
interact with an ambient medium (e.g., an ordinary interstellar
medium or plasma which is ejected from the predecessor of the
burster). Recently, the interaction between a relativistic magnetized wind
and an ambient medium was studied numerically (\cite{Smolsky96};
\cite{Usov98}), and it was shown that electrons of the ambient medium 
which are reflected from the wind front are accelerated up to the mean 
energy of reflected protons. In this paper we 
study both the spectrum of electrons accelerated 
at the wind front and their non-thermal radiation.

A plausible model of cosmological GRBs is discussed in \S~2. The main 
results of our numerical simulations of the interaction between a
relativistic strongly magnetized wind and an ambient medium are
presented in \S~3. The spectrum of electrons which are 
accelerated at the wind front and their non-thermal radiation
are considered in \S~4. Finally, our main conclusions are 
summarized and discussed in \S~5.
 
\section{Relativistic strongly magnetized winds from cosmological $\gamma$-ray 
bursters and their non-thermal radiation: a plausible scenario}
\label{model} 

The energy output of cosmological $\gamma$-ray bursters in 
$\gamma$-rays typically is $10^{51}-10^{53}$ ergs (\cite{Wickramasinghe93};
\cite{Tamblyn93}; \cite{Lipunov95}  ) and may be as high as 
$3\times 10^{53}$ ergs (\cite{Kulkarni98}) or even more (\cite{Kulkarni99}).
These estimates assume isotropic emission of GRBs.
Such a high energetics of cosmological $\gamma$-ray bursters
and a short time scale of $\gamma$-ray flux variability call for very
compact objects as sources of GRBs (\cite{Hartman95};
\cite{Piran98} and references therein). These objects may
be either millisecond pulsars which are arisen from accretion-induced
collapse of white dwarfs in close binaries (\cite{Usov92}) or differentially
rotating disk-like objects which are formed by the merger of a binary
consisting of two neutron stars (\cite{Eichler89}; \cite{Narayan92}).
Such very young fast-rotating compact objects have two possible sources
of energy which may be responsible for radiation of cosmological GRBs. 
These are the thermal energy of the compact objects and the kinetic energy
of their rotation. The thermal energy may be transformed into
$\gamma$-rays by means of the following sequence of processes (for review, 
see \cite{Piran98}): (1) Emission of neutrinos and cooling of the object;
(2) Absorption of neutrinos $(\nu_i +
\bar\nu_i \rightarrow e^+ + e^- )$ and formation of a fireball which mainly
consists of electrons and positrons;
(3) Expansion of the fireball and formation of a relativistic shell, $\Gamma_0
\gtrsim 10^2$; (4) Interaction of the shell with an external medium and
acceleration of electrons to very high energies; and (5) 
Generation of $\gamma$-rays by high-energy electrons. The maximum thermal
energy, $Q_{\rm th} ^{\rm max}$, of very young compact objects  
is high enough to explain the energy output
of cosmological GBBs, $Q_{\rm th} ^{\rm max}\simeq$ a few $\times 10^{53}$ ergs.
However, the fraction of the thermal energy that
is converted into the energy of the electron-positron
fireball and then into the kinetic energy of the relativistic shell is very small
and cannot be essentially more than $10^{-3}-10^{-2}$
(\cite{GDN}; \cite{Eichler89}; \cite{JR};
\cite{Piran98} and references therein). Moreover, the efficiency of
transformation of the kinetic energy of a relativistic shell into radiation 
cannot be more than $30-40$\% (\cite{be}). 
Hence, neutrino powered winds outflowing from compact objects
may be responsible for the radiation of cosmological GRBs only if they are 
well collimated, with opening angle about a few degrees or even less.
For both young neutron stars and post-merger objects, such a collimation of 
neutrino powered winds is very questionable (e.g., \cite{Woosley93};
\cite{Piran98} and references therein).

The rotational energy of the compact objects at the moment of their formation
may be comparable with the thermal energy, $Q_{\rm rot}
^{\rm max}\simeq  Q_{\rm th} ^{\rm max}$. The
efficiency of transformation of the rotational energy to
the energy of a relativistic strongly magnetized wind and then
to the energy of high-frequency radiation may be as high as
almost 100\% (see Usov 1994a,b; Blackman et al. 1996 and below).
For some time the theoretical expectation has been that
rotation powered neutron stars (pulsars)
should generate collimated outflows (e.g., \cite{Benford84};
\cite{Michel85}; \cite{Sulkanen90}). The Crab, Vela, PSR B1509-58
and possible PSR B1951+32 all show evidence that this is indeed
the case (\cite{Hester98}; \cite{Gaensler99} and references therein).
If the energy flux from the source of GRB 990123
in the direction to the Earth is only about ten
times more than the energy flux averaged over all directions,
the model of GRBs based on the rotation powered winds
can easily explain the energetics of
such an extremal event as GRB 990123
(\cite{Kulkarni99}). Such an anisotropy of emission
from the burst sources  doesn't contradict available data on GRBs
(e.g., \cite{PL98}). In the case of typical GRBs with the energy output of
$10^{51}-10^{53}$ ergs, this model can explain their energetics even if
the emission of GRBs is nearly isotropic.
Therefore, the rotational energy of compact
objects is a plausible source of energy for cosmological GRBs,
not the thermal energy.
 
In many papers (e.g., \cite{Usov92}; \cite{Thompson93}; 
Blackman et al. 1996; \cite{Kluzniak98}), it was argued that
the strength of the magnetic field $B_{_S}$ at the surface of compact objects 
may be as high as $\sim 10^{16}$ G or even more.  Such a strong magnetic field 
leads to both deceleration of the rotation of the fast-rotating compact object
on a time  scale of seconds and generation of a strongly
magnetized wind that flows away from the object at a
relativistic speed, $\Gamma_0\simeq 10^2-10^3$ (e.g., \cite{Usov94a}).
The outflowing wind is Poynting flux--dominated, i.e., $\sigma = L_{\pm}/
L_{_P}\ll 1$, where 

\begin{equation}
L_{_P}\simeq {2\over 3} {B_{_S}^2R^6\Omega^4\over c^3}\simeq 2\times
10^{52}\left({B_{_S}\over  10^{16}\,\,{\rm G}}
\right)^2\left({R\over 10^6\,\,{\rm cm}}\right)^6
\left({\Omega\over 10^4\,\,{\rm s}^{-1}}\right)^4\,\,
{\rm ergs\,\,s}^{-1}
\end{equation}

\noindent
is the luminosity of the compact object in the Poynting flux, $ L_{\pm}$ is its 
luminosity in both electron-positron pairs and radiation, $c$ is the speed of 
light, $R$ is the radius of the compact object
and $\Omega$ is its angular velocity; $R\sim 10^6$~cm and $\Omega\sim 
10^4$~s$^{-1}$ for both millisecond pulsars and post-merger objects. 
For compact objects with extremely strong magnetic fields, 
$B_{_S}\sim 10^{16}$~G, it is expected that $\sigma $ is $\sim 0.01-0.1$ 
(\cite{Usov94a}).

 A plausible magnetic topology for a relativistic magnetized 
wind outflowing from an oblique rotator ($\vartheta\neq 0$) with a 
nearly dipole magnetic field is shown in Figure~1, 
where $\vartheta$ is the angle between the rotational axis 
and the magnetic axis. Near the rotational poles, 
the wind field should be helical (e.g., \cite{Coroniti90}). 
This is because the magnetic flux originates in a single polar cap.
Near the rotational equator, the toroidal magnetic 
field of the wind should be striped and alternates in polarity 
on a scalelength of $\pi (c/\Omega )\sim 10^7$~cm. 
These magnetic stripes are separated by thin current sheets $(J_\theta)$. 
Off the equator, the magnetic flux in the toward and away stripes is 
unequal if $\vartheta\neq  \pi/ 2$. In other words,
in the striped region, the wind field is a superposition of a pure
helical field and a pure striped field with nearly equal magnetic
fluxes in adjacent stripes.

Since the luminosity of a $\gamma$-ray
burster in a relativistic magnetized wind drops in time, $L_{_P} 
\propto t^{-\beta}$, the wind structure at the moment  $t\gg\tau_{_\Omega}$ 
is similar to a shell with the radius $r\simeq ct$, where $\beta$ is a numerical index,
$1\leq \beta\leq 2$, and
$\tau_{_\Omega}\sim 10^{-2}-10^2$~s is the characteristic time 
of deceleration of the compact object rotation due to the action of 
the electromagnetic torque and the torque related to 
generation of gravitational radiation (\cite{Usov92}; \cite{Yi98}).
The thickness of the shell is $\sim c\tau_{_\Omega}$. 

The strength of the magnetic field at the front of the wind is about

\begin{equation}
B\simeq B_{_S}\frac {R^3}{r_{\rm lc}^2r}
\simeq  10^{15}\frac Rr\left(
\frac{B_{_S}}{ 10^{16}\,\rm{G}}\right) \left(\frac\Omega
{10^4\,\rm{s}^{-1}}\right)^2\rm{G},
\label{Bobj}
\end{equation}
 
\noindent
where $r_{\rm lc}=c/\Omega=3\times 10^6(\Omega /10^4\,$s$^{-1})$~cm
is the radius of the light cylinder. 

Non-thermal radiation from a relativistic strongly magnetized wind 
depends on whether the wind is striped in the direction of its outflow 
or not. If the wind is striped (see Fig.1), the magnetic field is more 
or less frozen in the outflowing plasma at the distance $r\lesssim r_f$
(\cite{Usov94a}; \cite{Blackman98}), where

\begin{equation}
r_f\simeq 2\times 10^{14} \sigma^{3/4}\left(
{B_{_ S}\over  10^{16}\,{\rm G}}\right)^{1/2}
\left({\Omega\over 10^4\,{\rm s}^{-1}}\right)^{1/2}\,\,
{\rm cm}\,.
\label{rnth}
\end{equation}

\noindent At $r > r_f$, the wind density is not sufficient 
to screen displacement currents, and the striped component of
the wind field is transformed into large-amplitude
electromagnetic waves (LAEMWs) due to development of
magneto-parametric instability (\cite{Usov75b},  1994a,b; Blackman 
et al. 1996; \cite{Melatos96}). [For a criterion for electromagnetic
waves to be considered as LAEMWs see \S~4.] The typical frequency 
of generated LAEMWs is equal to $\Omega$, and their amplitude is 
$\sim B$. Outflowing particles are accelerated in the field of LAEMWs 
to Lorentz factors of the order of $10^6$, and generate non-thermal
synchro--Compton radiation with the typical energy of photons
in the range of a few $\times (0.1-1)$~MeV (Usov 1994a,b; 
Blackman et al. 1996; \cite{Blackman98}). A long high-energy tail
of the $\gamma$-ray spectrum may exist up to $\sim 10^4$~MeV.
This is consistent with the observed spectra of GRBs 
(\cite{Band93}; \cite{Schaefer94}; \cite{FM95}). 

The radiative damping length for LAEMWs generated at $r\sim r_f$
is a few orders of magnitude less than $r_f$ (\cite{Usov94a}). Therefore,
at $r\gg r_f$ LAEMWs decay 
almost completely, and their energy is transferred to 
high-energy electrons and then to X-ray and $\gamma$-ray 
photons. It is worth noting that when the magnetic 
axis is perpendicular to the rotational axis,  $\vartheta = \pi /2$, the 
electromagnetic field of the Poynting flux--dominated wind is purely
striped just as vacuum magnetic dipole waves (\cite{Michel71}).
In this case, almost all energy of the wind is radiated in X-rays and
$\gamma$-rays at $r\sim r_f$ (\cite{Usov94a};  \cite{Blackman96}; 
\cite{Blackman98}), and the total energy output in 
hard photons per a GRB may be as high as $Q^{\rm max}_{\rm rot}
\simeq$ a few $\times 10^{53}$ ergs.

At $r\gg r_f$ and $\vartheta \neq \pi /2$, the magnetic field is helical 
everywhere in the  outflowing wind (see Fig.~1). Such a relativistic strongly 
magnetized wind expands more or less freely up to the distance 

\begin{equation}
r_{\rm{dec}}\simeq 5\times 10^{16}\left({\frac{Q_{\rm{kin}}}{10^{52}\,
\rm{ergs}}}\right)^{1/3}\left({\frac n{1\,\rm{cm}^{-3}}}\right)^{-1/3}\left(
{\frac{\Gamma_0}{10^2}}\right)^{-2/3}\rm{cm},
\label{rdec}
\end{equation}

\noindent at which 
deceleration of the wind due to its interaction with an ambient
medium becomes important (\cite{Rees92}), where
$n$ is the density of the ambient medium and $Q_{\rm{kin}}$
is the kinetic energy of the outflowing wind, $Q_{\rm{kin}}
\leq Q_{\rm rot}\leq Q^{\rm max}_{\rm rot}$ . 
Substituting $r_{\rm{dec}}$ for $r$ into equation (\ref{Bobj}), we
have the following estimate for the magnetic field at the wind
front at $r\sim r_{\rm {dec}}$:

\begin{equation}
B_{\rm{dec}}\simeq 2\times 10^4\left(\frac{B_s}{10^{16}\,\rm{G}}\right)\left(
\frac\Omega{10^4\,\rm{s}^{-1}}\right)^2\left(
\frac{Q_{\rm{kin}}}{10^{52}\,\rm{ergs}}\right)^{-1/3}\left(\frac
n{1\,\rm{cm}^{-3}}\right)^{1/3}\left({\frac{\Gamma_0}{10^2}}\right)
^{2/3}\rm{G}\,,
\label{Bdec}
\end{equation}
 
For typical parameters of cosmological $\gamma$-ray bursters, $B_{_S}
\simeq 10^{16}$~G, $\Omega\simeq 10^4$~s$^{-1}$, $Q_{\rm{kin}}\simeq
10^{52}-10^{53}$~ergs and $\Gamma_0\simeq 10^2-10^3$, if the ambient
medium is an ordinary interstellar gas, $n\sim 1
-10^2$~cm $^{-3}$, from equation (\ref{Bdec}) we
have $B_{\rm{dec}}\simeq 10^4- 4\times 10^5$~G. 

It is suggested by M\'{e}sz\'{a}ros and Rees (1992) that in the process 
of the wind -- ambient medium interaction at $r\sim r_{\rm{dec}}$, 
an essential part of the wind energy may be transferred to 
high-energy electrons and then to high-frequency (X-ray and
$\gamma$-ray) emission. This suggestion is confirmed by
our numerical simulations (see \cite{Smolsky96}; \cite{Usov98} and below).
Hence, in our scenario in a general case, when the rotational axis
and the magnetic axis are not aligned $(\vartheta \neq 0)$ or
perpendicular $(\vartheta \neq \pi /2)$, there are at least 
two regions where powerful non-thermal X-ray and $\gamma$-ray emission
of GRBs may be generated. The first 
region is at $r \sim r_f \sim 10^{13}-10^{14}$~cm, and the second 
one is at $r\sim r_{\rm dec}\sim 10^{16}- 10^{17}$~cm.  
Acceleration of electrons and their radiation in the first radiating
region at $r\sim r_f$ were considered in details earlier
(Usov 1994a,b; Blackman et al. 1996; Blackman \& Yi 1998). 
Below, we consider the spectrum of electrons
accelerated at $r\sim r_{\rm dec}$ and their radiation.

\section{Interaction of a relativistic strongly magnetized wind with 
an ambient medium}
\label{interaction} 

For consideration of the interaction between a
relativistic magnetized wind and an ambient medium, it is convenient
to switch to the co-moving frame of the outflowing plasma (the wind frame).
While changing the frame, the magnetic and electric fields in the wind
are reduced from $B$ and $E=B[1-(1/\Gamma_0^2)]^{1/2}
\simeq B$ in the frame of
the $\gamma$-ray burster to $B_0\simeq B/\Gamma_0$ and $E_0=0$
in the  wind frame. Using this and equation (\ref{Bdec}), for
typical parameters of cosmological $\gamma$-ray bursters (see
\S~2) we have $B_0\simeq 10^2-10^3$~G at $r\simeq r_{\rm dec}$.

In the wind frame, the problem of the
wind -- ambient medium interaction is identical to the problem of
collision between a wide relativistic beam of cold plasma and a region
with a strong magnetic field which is called a magnetic barrier.
Recently, the interaction of a wide relativistic plasma beam with a magnetic 
barrier was studied numerically (\cite{Smolsky96}; \cite{Usov98}).
In these studies, the following initial condition of the beam -- barrier 
system was assumed.  Initially, at $t=0$, the ultrarelativistic 
homogeneous neutral beam of protons and electrons 
(number densities $n_p=n_e\equiv n_0$) runs along the $x$ axis and 
impacts at the barrier, where $n_0$ is constant. The beam is infinite 
in the $y$ -- $z$ dimensions and semi-infinite in the $x$ dimension.
The  magnetic field of
the barrier ${\bf B}_0$ is uniform and transverse to the beam velocity,
${\bf B}_0=B_0{\bf \hat e}_z\Theta[x]$, where 
$B_0$ is constant and $\Theta [x]$ is the step function equal to
unity for $x>0$ and to zero for $x<0$.  At the front of the barrier, 
$x=0$, the surface current $J_y$ runs along the $y$ axis
to generate the jump of the magnetic field. The value of this 
current per unit length of the front across the current direction is 
$cB_0/4\pi$. A 1${1\over 2}$D time-dependent solution for the problem
of the beam -- barrier interaction was constructed, i.e.,
electromagnetic fields (${\bf E}=E_x{\bf \hat
  e}_x+E_y{\bf \hat e}_y$; ${\bf B}=B{\bf \hat e}_z$) and motion of 
the beam particles in the $x$ -- $y$ plane were found. The structure of 
the fields and motion of the beam particles were treated self-consistently
except the external current $J_y$ which was fixed in 
our simulations. 

The main results of our simulations are the following
(\cite{Smolsky96}; \cite{Usov98}).

1. When the energy densities of the beam and the magnetic field,
${\bf B}_0$, of the barrier are comparable, 

\begin{equation}
\alpha = 8\pi n_0m_pc^2(\Gamma_0-1)/B^2_0\sim 1\,, 
\label{alpha}
\end{equation}

\noindent 
where $m_p$ is the proton mass, the process
of the beam -- barrier interaction is strongly nonstationary, and the 
density of protons after their reflection from the barrier
is strongly non-uniform. The ratio of the maximum density of
reflected protons and their minimum density is $\sim 10$.
 
2. At $\alpha > \alpha _{\rm cr}\simeq 0.4$, 
the depth of the beam particle penetration into the barrier 
increases in time, $x_{\rm pen}\simeq v_{\rm pen} t$, where 
$v_{\rm pen}$ is the mean velocity of the penetration
into the barrier. The value of $v_{\rm pen}$ is subrelativistic and 
varies from zero (no penetration)  at $\alpha \leq \alpha_{\rm cr}$ to
$0.17c$ at $\alpha =1$ and then to $0.32 c$ at $\alpha =2$.
At $\alpha > \alpha _{\rm cr}$,
the magnetic field of the barrier at the moment $t$ roughly
is $B(t)\simeq B_0\Theta[x-x_{\rm {pen}}(t)]$ (see Fig.~2).
In other words,  the front of the beam -- barrier interaction
is displaced into the barrier with the velocity $v_{\rm pen}$.
For $\alpha > \alpha _{\rm cr}$, our consideration of the beam -- barrier 
interaction in the vicinity of the new front, $x\simeq x_{\rm {pen}}$, 
is {\it completely self-consistent}, and no 
simplifying assumptions besides geometrical ones are exploited.

3. At the front of the barrier, $x\simeq x_{\rm {pen}}(t)$, the surface 
current varies in time because of strong nonstationarity 
of the beam -- barrier interaction at $\alpha\sim 1$,
and low-frequency electromagnetic waves are generated (Fig.~2).
The typical frequency of these waves is about the proton 
gyrofrequency $\omega_{Bp}=eB_0/m_pc\Gamma_0$
in the field of the barrier $B_0$. The wave amplitude 
$ B_w$ can reach $\sim 0.2 B_0$.

4. At $\alpha \sim 1$, strong electric fields are generated in the vicinity 
of the front of the barrier, $x\simeq x_{\rm {pen}}(t)$, and electrons of 
the beam are accelerated in these fields up to the mean energy 
of protons, i.e. up to $\sim m_pc^2 \Gamma_0$ (see Fig.~3). 
At $\alpha_{\rm cr} < \alpha \lesssim 1$,
the mean Lorentz factor of outflowing high-energy electrons 
after their reflection and acceleration at the barrier front
depends on $\alpha$ and is 

\begin{equation}
\langle\Gamma_e^{\rm {out}}\rangle \simeq 0.2
\left(\frac{m_p}{m_e}\right)\Gamma_0
\label{Gammamean} 
\end{equation}

\noindent 
within a factor of 2. The total energy of accelerated electrons is
about 20\% of the energy in outflowing protons which are
reflected from the magnetic barrier. 

5. At $\alpha_{\rm cr} < \alpha \lesssim 1$, the mean Lorentz 
factor of protons reflected from the barrier is
$\langle \Gamma_p^{\rm{out}}\rangle \simeq (0.7\pm 0.1)\Gamma_0$,
i.e. the process of the beam proton reflection from the barrier
is non-elastic, and about  30\% of the initial kinetic energy 
of the beam protons is lost in
this collision. The energy that is lost by the beam protons 
is transferred to high-energy electrons and 
low-frequency electromagnetic waves. 
Typically, the energy in these waves is a few
times smaller than the energy in high-energy electrons. 

In the burster frame, a magnetized wind flows away
from the burster at relativistic speeds and collides with an ambient
medium. In the process of such a collision, the outflowing wind loses its 
energy. From the listed results of our simulations 
of the beam -- barrier collision
(\cite{Smolsky96}; \cite{Usov98}), it follows that at $r\sim r_{\rm dec}$,
where $\alpha$ is $\sim 1$, about 70\% of the energy losses of
the wind are transferred to protons of the ambient medium
which are reflected from the wind front. The mean
energy of reflected protons is about $m_pc^2\Gamma_0^2$. The other
30\% of the wind energy losses are distributed
between high-energy electrons and low-frequency 
electromagnetic waves. As a rule, the total energy in accelerated electrons
is a few times more than the total energy in low-frequency waves.

\section{Non-thermal radiation from the region of the wind --
ambient medium interaction}

High-energy electrons accelerated at the wind front generate
non-thermal radiation while they move in both the magnetic field of
the wind and the electromagnetic fields of low-frequency
waves that are produced by the non-stationary currents at the 
wind front.  

\subsection{Synchrotron radiation from the wind front}

In our simulations of the beam -- magnetic barrier interaction 
(\cite{Smolsky96}; \cite{Usov98}), the examined space-time domain is

\begin{equation}
x_{\rm min}<x<x_{\rm max},\,\,\,\,\,\,\,\,\,\,0<t<t_{\rm max},
\label{xt}
\end{equation}

\noindent
where $x_{\rm max}$ and $-x_{\rm min}$ are equal to a few $\times (1-10)
(c/\omega_{Bp})$, $t_{\rm max}$ is equal to a few $\times (1-10)T_p$, and
$(c/\omega_{Bp})$ and $T_p=2\pi / \omega_
{Bp}$ are the proton gyroradius and gyroperiod in the magnetic field of 
the barrier $B_0$, respectively. Non-thermal radiation of high-energy 
electrons from the examined space domain was calculated for the beam 
and barrier parameters which are relevant to cosmological GRBs 
(\cite{Smolsky96}; \cite{Usov98}).  Figure~4 shows the intensity of 
radiation as a function of time $t$ for a simulation with $B_0=300$~G,
$\Gamma_0=300$, $\alpha =2/3$, $x_{\rm min}=-5(c/\omega_{Bp})$
and $x_{\rm max}=30(c/\omega_{Bp})$. In all our simulations, the bulk of 
calculated radiation is generated via synchrotron mechanism
in a compact vicinity, $x_{\rm pen}- 2 (c/\omega_{Bp}) <x
< x_{\rm pen}$, of the barrier front where both the strength of the magnetic
field is of the order of $B_0$ and the mean energy of accelerated 
electrons is extremely high. Radiation of high-energy
electrons in the fields of low-frequency waves is negligible
($\sim 1$\% or less) because both the fields of these waves are about
an order of magnitude smaller than $B_0$ (see below) and
the length of the examined space domain (\ref{xt})
is restricted for computational reasons.  

At $\alpha\sim 1$, the mean energy of
synchrotron photons generated at the front of the barrier is

\begin{equation}
\langle \varepsilon _\gamma \rangle \simeq 0.1\left( {\frac{\Gamma
_0}{10^2}} \right) ^2\left( {\frac{B_0}{10^3\,\rm{G}}}\right) \;\;
\rm{MeV}\,. 
\label{eps}
\end{equation}
 
\noindent 
The average fraction of the kinetic energy of the beam that is
radiated in these photons is 
 
\begin{equation}
\xi _\gamma \equiv \frac{\left\langle \Phi _\gamma \right\rangle }{
n_0m_pc^3\Gamma _0}
\simeq 10^{-3}\left( {\frac{\Gamma _0}{10^2}}\right)
^2\left( {\frac{B_0}{10^3\,\rm{G}}}\right)\,,  
\label{ksinum}
\end{equation}
 
\noindent where $\left\langle\Phi_\gamma \right\rangle$
is the average synchrotron luminosity of
high-energy electrons per unit area of the barrier front. 

In the burster frame, the characteristic energy, $\langle \tilde{\varepsilon}_
\gamma \rangle$, of synchrotron photons generated in the vicinity of the 
wind front increases due to the Doppler effect. Taking this
into account and using equation (\ref{eps}), we have 

\begin{equation}
\langle \tilde{\varepsilon}_\gamma \rangle \simeq
10\left( {\frac{\Gamma _0 }{10^2}}\right) ^2\left(
{\frac{B_{\rm{dec}}}{10^5\,\rm{G}}}\right) \;\;\rm{ MeV}\,.
\label{epsLab}
\end{equation}

\noindent
where $B_{\rm{dec}}$ is determined in equation (\ref{Bdec}).
The fraction of the wind energy which is
transferred to radiation at the wind front does not depend on the frame
where it is estimated, and in the burster frame it is equal to

\begin{equation}
\tilde{\xi}_\gamma = \xi_\gamma\simeq  10^{-3}\left( \frac{\Gamma
_0}{10^2} \right) \left( \frac{B_{\rm{dec}}}{10^5\, \rm{G}}\right) \,,
\label{ksiLab} 
\end{equation}
 
\noindent 
For typical parameters of relativistic magnetized winds which 
are relevant to cosmological $\gamma$-ray bursters,
$\Gamma_0\simeq 10^2-10^3$ and $B_{\rm{dec}}\simeq 10^5$~G
(see \S~2), from equations (\ref{epsLab}) and  (\ref{ksiLab})
we have $\langle \tilde{\varepsilon}_\gamma \rangle\simeq 10 - 
10^3$~MeV and $\tilde{\xi}\simeq 10^{-2}-10^{-3}$, i.e., the synchrotron
radiation that is generated at wind front is in the EGRET range. 

The main part of X-ray and $\gamma$-ray emission of
detected GRBs is in the BATSE range, from a few $\times 10$ keV to a few
MeV (e.g., \cite{FM95}). Synchrotron radiation from the wind front is either 
too hard or too weak to explain this emission irrespective of $B_{\rm dec}$. 
Indeed, equations (\ref{epsLab}) and (\ref{ksiLab}) yield
 
\begin{equation}
\langle \tilde{\varepsilon}_\gamma \rangle \simeq 10^2\left(
{\frac{\Gamma _0}{10^2}}\right) \left( {\frac{\tilde{\xi}_\gamma
}{10^{-2}}} \right)\;\;\;\rm{MeV}.  
\label{epsLabksiLab}
\end{equation}

\noindent In our model, the energy of rotation powered winds
which are responsible for cosmological GRBs cannot be significantly 
more than $10^{53}$~ergs. To explain the energy output of $\sim 10^{51}-
10^{53}$~ergs per GRB (see \S~2), the  efficiency of transformation
of the wind energy into the energy of non-thermal radiation has to be
about or more than $ 1\%$, $\tilde{\xi}_\gamma \gtrsim 10^{-2}$,
and for some GRBs it may be as high as 100\%, $\tilde{\xi}_\gamma 
\simeq 1$. Taking into 
account that for cosmological GRBs the value of $\Gamma _0$ is 
$\gtrsim 10^2 $ (e.g., \cite{Fenimore93}; \cite{Harding97}), 
for $\tilde{\xi}_\gamma \gtrsim 10^{-2}$ from equation 
(\ref{epsLabksiLab}) it follows that the mean energy of synchrotron
photons generated at the wind front is very high, $\langle
\tilde{\varepsilon}_\gamma\rangle\gtrsim 100$~MeV. This is because
the bulk of these photons is generated in the thin vicinity 
of the wind front where there are both a strong magnetic field, 
$B\simeq B_0$, and high-energy electrons. The mean time that
high-energy electrons spend in this vicinity and generate 
synchrotron radiation is very short, $\sim$ a few $\times T_p\sim
10^{-6}(B_{\rm dec}/10^5\,\,{\rm G})^{-1}(\Gamma_0/10^2)$~s.
To get a high efficiency of synchrotron radiation at the wind front, 
$\tilde{\xi}_\gamma > 10^{-2}$, it is necessary to assume that 
the magnetic field $B_{\rm dec}$ is about its maximum value 
(see \S~2). From equation (\ref{epsLab}), we can see that
in this case the mean energy of synchrotron photons is 
of the order of or higher than $10^2$~MeV.
 
For a typical value of $B_{\rm dec}$, that is $\sim 10^5$~G (see \S~2),
and $\Gamma_0\sim 10^2-10^3$, from equation (\ref{ksiLab})
we have $\tilde \xi_\gamma\sim 10^{-3}-10^{-2}\ll 1$.
Hence, high-energy electrons are accelerated at 
the wind front and injected into the region ahead of the front
practically without energy losses. In our model,
these high-energy electrons are the best candidates to be
responsible for the X-ray and $\gamma$-ray emission of GRBs in the 
BATSE range. Before considering
high-frequency radiation from the region ahead of the wind front, 
we discuss briefly some general properties of radiation of 
high-energy electrons in the fields of electromagnetic waves.

\subsection{LAEMWs and synchro-Compton radiation of electrons}

The motion of electrons and their radiation in the fields of
electromagnetic waves is characterized by 
the following dimensionless Lorentz-invariant parameter
(e.g., \cite{BT74}):

\begin{equation}
\eta ={e B_w\over m_ec\omega}\,,
\label{eta}
\end{equation}

\noindent
where $ B_w$ is the amplitude of the waves and $\omega$ is their 
frequency. At $\eta \ll 1$, electrons radiate via Compton
scattering. In this case, the typical energy of 
photons $\langle{\varepsilon}_c\rangle$ after scattering 
depends on the wave frequency, 
$\langle{\varepsilon}_c\rangle\propto \omega \Gamma_e^2$, and
does not depend on the wave amplitude (e.g., \cite{BT74}).
 
Electromagnetic waves with $\eta \gg 1$ are called 
large-amplitude electromagnetic waves 
(LAEMWs). Radiation of electrons in the fields of LAEMWs
is called synchro--Compton radiation. In the most simple
case which is interesting for us (see below) when the bulk of
radiating electrons has such a high velocity $v_\perp=c\sin \psi$
transverse to the wave vector that 

\begin{equation}
(v_\perp /c)\Gamma_e = \Gamma_e\sin\psi \gtrsim \eta\,,
\label{psi}
\end{equation}

\noindent synchro--Compton radiation of these electrons
closely resembles synchrotron radiation (\cite{BT74}),
where $\psi$ is the angle between 
the electron velocity and the wave vector of LAEMWs.
Indeed, it is well known that electromagnetic radiation of 
relativistic electrons is concentrated in the direction
of the particle's velocity within a narrow cone of angle 
$\Delta \varphi\simeq 1/\Gamma_e \ll 1$
(e.g., \cite{Rybicki79}). For nearly transverse motion of 
electrons in a magnetic field $B$, the formation length of radiation 
is about $\Delta l \simeq 2 R_{Be}\Delta \varphi\simeq 
2m_ec^2/eB$, where $R_{Be}=m_ec^2\Gamma_e/eB$ 
is the electron gyroradius.
For synchro--Compton radiation, $\eta\gg 1$,
the formation length is much smaller than the wavelength, 
$\lambda =2\pi c/\omega$, of LAEMWs :
 
\begin{equation}
\Delta l\simeq \left({m_ec \omega \over \pi eB}\right)\lambda 
\simeq {\lambda \over \pi\eta} \ll \lambda\,,
\label{Deltal}
\end{equation}

\noindent
and the electromagnetic fields of LAEMWs 
may be considered as homogeneous and static. 
Therefore, synchro--Compton radiation  
is similar to synchrotron radiation in a magnetic field 
which is equal to the local field of LAEMWs. 
[For details on synchro--Compton radiation see 
(\cite {GO71}; \cite{BT74}).]  

The mean frequency of  synchro-Compton radiation of 
high-energy electrons with the Lorentz factor $\Gamma_e$ 
is (e.g., \cite{BT74})

\begin{equation}
\nu_{sc}\simeq {e\langle B_w\rangle \Gamma^2_e
\over 2\pi mc}\simeq 3\times 10^6\langle B_w\rangle
\Gamma_e^2\,\,\,\,{\rm Hz}\,,
\label{nu}
\end{equation}

\noindent where $\langle B_w\rangle$  is the mean field of LAEMWs. 

\subsection{LAEMWs generated at the wind front}

At $\alpha\sim 1$, low-frequency electromagnetic waves are 
generated at the wind front and propagate in both 
directions from the front (see \S~3 and Figs~2 and 5 -- 7). 
These waves are non-monochromatic. Figure~8 shows a 
typical spectrum of low-frequency waves in the wind frame. 
This spectrum has a maximum at the frequency $\omega_{\rm max}$ 
which is a few times higher than the proton
gyrofrequency $\omega_{Bp}=eB_0 /m_pc\Gamma_0$
in the field of $B_0$. 

At high frequencies, $\omega > \omega_{\rm max}$,
the spectrum of low-frequency waves may be fitted by a power law:

\begin{equation}
|B(\omega )|^2\propto 
\omega ^{-\beta}\,,
\label{Bomega}
\end{equation}

\noindent
where $\beta\simeq 1.6$. 

In the region ahead of the wind front,
the mean field of low-frequency waves depends on $\alpha$, 
and in the wind frame, for $\alpha_{\rm cr}\lesssim \alpha \lesssim 1$  
this field is

\begin{equation}
\langle B_w\rangle =(\langle B_z\rangle^2 + 
\langle E_y\rangle^2)^{1/2}\simeq 0.1 B_0 \simeq 0.1 B_{\rm dec}
/\Gamma_0
\label{avBw}
\end{equation}

\noindent
within a factor of 2 or so. 

At $\alpha > 1$, the intensity of low-frequency waves 
is suppressed in the region ahead of the wind front (see Fig.~6) 
because the typical frequency, $\omega\simeq \omega_{\rm max}$,
of these waves is smaller than the 
relativistic plasma frequency of the ambient medium, $\omega_p=
(4\pi n_0e^2/m_e\langle\Gamma_e\rangle )^{1/2}$
(\cite{Gallant92}). In this case, only a high-frequency component of 
low-frequency waves with the frequency 
$\omega \gtrsim \omega_p\simeq (5\alpha )^{1/2}\omega_{Bp}$ 
propagates far ahead of the wind front.

For typical parameters of low-frequency waves generated at the wind 
front, $B_w\simeq 0.1 B_0$ and $\omega\simeq \omega_{\rm max}\simeq 3
 \omega_{Bp}\simeq 3 (eB_0/m_pc\Gamma_0)$,
from equation (\ref{eta}) we have 

\begin{equation}
\eta \simeq {B_w\over B_0}{m_p\over m_e}\Gamma_0
\simeq 50\Gamma_0\gg 1\,.
\label{eta1}
\end{equation}

\noindent 
Hence, these low-frequency waves are LAEMWs. 

\subsection{High-energy electrons accelerated at the wind front}

At $r\sim r_{\rm dec}$, where $\alpha$ is $\sim 1$, about 20\%
of the energy of a relativistic strongly magnetized wind
is transferred  to electrons of an ambient medium which are reflected 
from the wind front and accelerated to extremely high energies
(see \S~3 and references  therein). In the wind frame, the spectrum 
of high-energy electrons in the region ahead of the wind front 
may be fitted by a two-dimensional relativistic Maxwellian 

\begin{equation}
{dn_e \over d\Gamma_e}\propto \Gamma_e\exp \left({-
{m_ec^2\Gamma_e\over kT}}\right)
\label{dne}
\end{equation}

\noindent
with a relativistic temperature $T=m_ec^2\Gamma_{_T} /k$, where 
$\Gamma_{_T}\simeq 240 \Gamma_0$ (see Fig. 9). The fact that the 
energy distribution of accelerated electrons at $\alpha\sim 1$  
is close to 
a relativistic Maxwellian is quite natural because at $\alpha 
\gtrsim \alpha_{\rm cr}$ the trajectories of particles in the
front vicinity are fully chaotic. 
The thermal distribution of high energy electrons does not
come as a result of interparticle collisions, since the ambient
medium is collisionless and no artificial viscosity is included in
the simulation code (\cite{Smolsky96}). Thermalization of the
electron distribution is purely a result of collisionless interactions
between particles and electromagnetic oscillations generated at the wind front.

At $\Gamma_e\leq 700\Gamma_0$, the fit of
the electron spectrum by a two-dimensional 
relativistic Maxwellian (\ref{dne}) with $\Gamma_{_T}
\simeq 200\Gamma_0$ is rather accurate (see Fig.~9). 
A small excess of electrons
with Lorentz factors $\Gamma_e > 700 \Gamma_0$ may be interpreted as 
a high-energy tail. Such a tail may result, for example,  
from multiple acceleration of high-energy electrons at the wind front.
Figure 10 shows the angular distribution of high-energy electrons in the 
region far ahead of the wind front. From Figure~10,
we can see that this distribution is anisotropic. The mean angle
between the velocity of outflowing electrons and the normal to 
the wind front is $\langle \psi \rangle\simeq 1/3$ radian.

\subsection{Synchro-Compton radiation of high-energy electrons
from the region ahead of the wind front}

High-energy electrons with a nearly Maxwellian spectrum are injected 
into the region ahead of the wind front and radiate in the fields of 
LAEMWs via synchro-Compton mechanism. The mean angle between the 
velocity of these electrons and the wave vector of LAEMWs is
$\langle\psi \rangle \simeq  1/3$ radian. Using this and equations 
(\ref{Gammamean}) and (\ref{eta1}), we can see that the condition 
(\ref{psi}) is satisfied, and, therefore, in the region ahead of the wind front
synchro-Compton radiation of high-energy electrons closely
resembles synchrotron radiation. To model the
spectrum of synchro-Compton radiation, we replace the fields 
of LAEMWs $B_z(t,x)$ and $E_y(t,x)$ by a constant magnetic field
which is equal to the mean field of LAEMWs $\langle B_w\rangle$.
The energy losses of electrons in such a magnetic field are 
governed by (\cite{Landau71})

\begin{equation}
  \frac{d\Gamma_e}{dt}=-\chi (\Gamma_e^2 -1)\,,
  \label{SynchLoss}
\end{equation}

\noindent where $\chi = 2e^4\langle B_w\rangle^2/3m_e^3c^5$.

In our approximation, the evolution of the spectrum
of high-energy electrons in the region ahead of the 
wind front may be found from the following
equation (e.g., \cite{Pacholczyk69})

\begin{equation}
{\partial f(\Gamma_e, t)\over \partial t}=
{\partial \over \partial\Gamma_e}[\chi \Gamma_e^2
 f(\Gamma_e, t)] + \dot N _ef_{\rm bar}(\Gamma_e)\,,
\label{par}
\end{equation}

\noindent where $f(\Gamma_e, t)$ is the distribution function
of high-energy electrons in the region ahead of the 
wind front per unit area of the front at the moment $t$, 
$\dot N_e\simeq n_0c$ is the rate of production of high-energy 
electrons per unit area of the front, and 
$f_{\rm bar}=(\Gamma_e/\Gamma_{_T}^2)\exp (-\Gamma_e/
\Gamma_{_T})$ is the average spectrum of high-energy electrons
which are injected into the region ahead of 
the front. The function  $f(\Gamma_e, t)$ is 
normalized to the total number of high-energy electrons
per unit area of the front $N_e$, while the function $f_{\rm bar}$
is normalized to unity:

\begin{equation}
\int_1^{\infty}f(\Gamma_e, t)\,d\Gamma_e=N_e\,\,\,\,\,\,\,
{\rm and }\,\,\,\,\,\,\,\int_1^{\infty}f_{\rm bar}\,d\Gamma_e=1\,.
\label{norm}
\end{equation}

\noindent
For simplicity, we disregard the angular anisotropy of the electron 
distribution function.

Under the mentioned assumptions, in the frame of the wind front
the differential proper intensity of synchro-Compton radiation 
from the region ahead of the wind front
is (e.g., \cite{Pacholczyk69}; \cite{Rybicki79})

\begin{equation}
I_\nu (t)= \int_1^{\infty}f(\Gamma_e, t)i_\nu d\Gamma_e\,,
\label{Inu}
\end{equation}

\noindent where 

\begin{equation}
i_\nu =\frac{\sqrt{3}e^3\langle B_w\rangle}{m_ec^2}
\frac \nu {\nu _c}\int_{\nu /\nu
_c}^\infty K_{5/3}(\eta )d\eta \,,  
\label{specti}
\end{equation}
 
\noindent 
is the spectrum of synchrotron radiation generated by
a single relativistic electron in a uniform magnetic field $\langle B_w\rangle$, 
$K_{5/3}$ is the modified Bessel functions of 5/3 order and
 
\begin{equation}
\nu _c={\frac{3e\langle B_w\rangle\Gamma^2_e}{4\pi m_ec}}
\label{nuTyp}
\end{equation}

\noindent
is the typical frequency of synchrotron radiation.

The observed spectral flux $F_\nu (t)$ (in units erg s$^{-1}$
cm$^{-2}$ erg$^{-1}$) can be obtained from the proper
intensity $I_\nu (t)$ dividing by the square of the burster
distance $d$, and by taking into account both the effects of
relativistic beaming and cosmological effects
(e.g., \cite{Tavani96a}; \cite{Dermer98}):

\begin{equation}
F_\nu (t)={D^3(1+z)\over 4\pi d^2}I_{\nu'}(t)\,,
\label{Fnu}
\end{equation}

\noindent where $D$ is the relativistic Doppler factor, 
$D\simeq 2\Gamma_0$, and $z$ is the cosmological redshift.
The observed frequency $\nu$ depends on the emitted frequency
$\nu'$ in the wind frame as $\nu=[D/(1+z)]\nu'\simeq [2\Gamma_0/
(1+z)]\nu'$. 

Equations (\ref{SynchLoss}) -- (\ref{Fnu}) were integrated
numerically. For typical parameters of cosmological $\gamma$-ray
bursters, $\Gamma_0 = 150$, $B_0= 300$~G, $\langle B_w\rangle =
0.1 B_0$, $\Gamma_T = 200\Gamma_0 = 3\times 10^4$ and $z=1$, 
Figure~11 shows the observed spectrum of synchro-Compton 
radiation from the region ahead of the wind front.
For this radiation the characteristic 
energy of photons is in the BATSE range  (e.g.,  \cite{Band93};  
\cite{Schaefer94};  \cite{FM95}; \cite{Preece96}; \cite{Schaefer98}).
The spectrum  of synchro-Compton radiation displays a continuous
hard to soft evolution, in agreement with observational data on
GRBs (\cite{Bhat94}). 

\section{Conclusions and discussion}
\label{discussion}

In a relativistic strongly magnetized wind outflowing from a 
fast rotating compact object like a millisecond pulsar
with the surface magnetic field $B_{_S}\sim 10^{15}-10^{16}$~G, 
there are at least two regions where extremely 
powerful non-thermal emission in hard (X-ray and $\gamma$-ray)
photons may be  generated (see Fig. 1). The first radiating region 
is at the distance $r_f\sim 10^{13}-10^{14}$~cm
from the compact object. In this region, the striped component of
the wind field is transformed into LAEMWs.
Acceleration of electrons in the fields of LAEMWs and generation of
non-thermal radiation at $r\sim r_f$ were considered in
(Usov 1994a,b; Blackman et al. 1996; Blackman \& Yi 1998). The second 
radiating region is at $r\sim r_{\rm dec}\sim 10^{16}-10^{17}$~cm
where deceleration of the wind due to its interaction with an
ambient medium becomes important. We have used one-and-a-half
dimensional particle-in-cell plasma simulations to study both 
the interaction of a relativistic, strongly magnetized wind with an 
ambient medium at  $r\sim r_{\rm dec}$ and non-thermal radiation 
from the region of this interaction (\cite{Smolsky96}; \cite{Usov98} 
and the present paper).  One of the main results of this study
is that we have confirmed the idea of M\'esz\'aros and Rees (1992)
that the wind -- ambient medium interaction may be responsible for
high-frequency (X-ray and $\gamma$-ray) emission of cosmological
GRBs. We have shown that at $r\sim r_{\rm dec}$, where in the wind frame
the magnetic pressure of the wind is comparable to  the dynamical pressure
of the ambient medium, $\alpha \sim 1$,
an essential part (about 20\%) of the wind energy may be transferred to
high-energy electrons and then to high-frequency emission. 
In this paper, it is shown that in the wind frame the spectrum of
electrons which are accelerated at the wind front and move ahead
of the front is close to a two-dimensional relativistic
Maxwellian with the temperature $T=m_ec^2
\Gamma_{_T}/k$, where $\Gamma_{_T}$ is equal to $200\Gamma_0$
with the accuracy of $\sim 20$\%. Our simulations point to
the existence of a high-energy tail of accelerated electrons with
a Lorentz factor of more than $\sim 700 \Gamma_0$.
It was shown in our early papers (\cite{Smolsky96}; \cite{Usov98})
that the process of the wind -- ambient
medium interaction is strongly nonstationary, and LAEMWs are
generated by the oscillating currents at the wind front and
propagate away from the front. In the wind frame,
the mean field of LAEMWs in the region ahead of the wind front is 
$\langle  B_w\rangle\simeq 0.1 B_{\rm dec}/\Gamma_0 \sim 10-10^2$~G.
High-energy electrons that are accelerated at the wind front
and injected into the region ahead of the front 
generate synchro-Compton radiation in the fields of LAEMWs. This 
radiation closely resembles synchrotron radiation in a uniform magnetic
field with the strength of $\langle  B_w\rangle$. In the burster frame, for
typical parameters of relativistic strongly magnetized
winds outflowing from millisecond pulsars with extremely strong
magnetic fields, $B_{_S}\simeq 10^{15}-10^{16}$~G, 
the maximum of synchro-Compton  radiation generated 
in the region ahead of the wind front at $r\sim r_{\rm dec}$
is in the BATSE range, from $\sim 10^2$~keV
to a few MeV. Radiation which is generated at the
wind front may be responsible for high-energy $\gamma $-rays,
$\varepsilon_\gamma>$ a few ten MeV, which are observed in the 
spectra of some GRBs (\cite{Hurley94}). The energy 
flux of GRBs in such high-energy $\gamma $-rays may be more 
or less comparable with the energy flux in the BATSE
range if $\Gamma_0B_{\rm dec} \gtrsim 3\times 10^8$~G. 
Besides that, high-energy $\gamma$-rays may be generated efficiently
via inverse Compton scattering (e.g., \cite{Jones79};
\cite{Meszaros94}). In the burster frame, the spectrum of  
high-energy $\gamma$-rays generated at $r\sim r_{\rm dec}$
may, in principle, extend up to  the maximum energy of 
accelerated electrons which is about the maximum energy of 
protons reflected from the wind front, $\varepsilon _\gamma
^{\rm max}\sim m_p\Gamma_0^2\sim 10^{13}(\Gamma_0/10^2)^2$~eV. 
In connection with this, it is worth noting that the Tibet air
shower array (\cite{Amenomori96}) and HEGRA AIROBICC Cherenkov 
array (\cite{Padilla98}) have independently reported significant 
excesses of $\gamma$-rays at energies of $\sim 10^{13}$~eV
coincident with some GRBs both in direction and burst time.
The statistical significance of these results
was estimated to be about $6\,\sigma$.
[For details on very high energy $\gamma$-rays from cosmological
$\gamma$-ray bursters see (\cite{Totani98}, 1999).]

To fit the observed spectra of GRBs,
for accelerated electrons it was usually taken either a 
three-dimensional relativistic Maxwellian distribution (Katz 1994a,b)
or a sum of a three-dimensional relativistic Maxwellian distribution 
and a suprathermal power-law tail (Tavani 1996a,b).
The thermal character of the electron distribution (or its part) is consistent 
with our results. In our simulations, we observe a two-dimensional relativistic 
Maxwellian distribution of accelerated electrons.  However, the difference
between 2D and 3D Maxwellian distributions does not
change the calculated spectrum of synchrotron radiation
significantly (\cite{Jackson75}; \cite {Jones79}; \cite{Rybicki79}).

Our solution for the wind -- ambient medium collision is completely 
self-consistent only if $\alpha$ is  $\gtrsim \alpha_{\rm cr}\simeq 0.4$ 
(\cite{Usov98}). Such values of $\alpha$ are
most interesting for cosmological GRBs because at $r\simeq 
r_{\rm dec}$ when deceleration of a relativistic wind 
due to its interaction with the ambient medium becomes important
the value of $\alpha$ is about unity. At $\alpha \ < \alpha_{\rm cr}$, both
the structure of the wind front and the process of the wind -- 
ambient medium interaction are not studied well enough. 
For this case, we have constructed a solution with a strong surface 
current at the wind front (\cite{Smolsky96}; \cite{Usov98}). 
In this solution, the surface current is considered as an
external one and fixed. Maybe, in the process of interaction of
a relativistic strongly magnetized wind with an ambient medium
at $\alpha \ < \alpha_{\rm cr}$ 
the structure of the wind front is self-regulated so that 
a required current runs along the front.

A situation similar, in some respects, to our problem was considered in
the works of Hoshino et. al (1992) and Gallant \& Arons (1994), where
collisionless shocks rather than the plasma beam -- magnetic barrier
collision were examined. In these studies,  acceleration of light particles,
electrons and positrons,
was observed near the shock (\cite{Hoshino92}). The level of
electron acceleration reported by Hoshino et. al (1992) is compatible
with our results.  Indeed, the energy that is transferred from 
protons to electrons depends on their mass ratio, $m_p/m_e$.
For computational reasons, the
mass ratio used by Hoshino et. al (1992) was small,
$m_p/m_e=20$. In our simulations, we use the realistic mass
ratio, $m_p/m_e=1836$. If we substitute $m_p/m_e=20$ into equation
(\ref{Gammamean}), we obtain the mean Lorentz factor of
accelerated electrons that coincides with the results of Hoshino et
al. (1992) within uncertainties of their and our calculations.  
The energy spectrum of accelerated electrons in the simulations
of Hoshino et. al (1992) is close to a relativistic Maxwellian too.

In our scenario, the rotational energy of compact objects (millisecond 
pulsars or post-merger objects) is a source of energy for emission of 
cosmological $\gamma$-ray bursters. This energy may be 
as high as a few $\times10^{53}$ ergs.  If radiation of the
gravitational waves by the compact object is negligible, the rotational 
energy is transformed to the  energy of a relativistic Poynting 
flux-dominated wind with the efficiency of $\sim 100\%$. 
In the case when the angle $\vartheta$ between the rotational and
magnetic axes of the compact object is about $\pi /2$,
almost all the energy of the wind is radiated in X-rays and
$\gamma$-rays at $r\sim r_f$ (\cite{Usov94a}; Blackman et al. 1996; 
\cite{Blackman98}). The maximum energy which may be
radiated in hard photons at $r\sim r_{\rm dec}$ is only a few times
smaller than that at $r\sim r_f$. Hence, the total energy 
output in hard photons per GRB may be as high as $\sim 10^{53}$ 
ergs or even a few times more. This energy is sufficient for the 
explanation of the energetics of cosmological GRBs.

At $\vartheta\simeq \pi /2$, when the bulk of the 
wind energy is transferred into $\gamma$-rays at $r\sim r_f$,
the residual energy of the wind at $r\gg r_f$ may be very small. 
In this case, afterglows which are generated at $r\gtrsim r_{\rm dec}$ and
accompany GRBs are weak irrespective of that the GRBs themselves 
may be quite strong. This may explain the fact that X-ray, optical and
radio afterglows have been observed in some strong GRBs but not 
in others (e.g., \cite{Piran98} and references therein).

For interpretation of data on cosmological GRBs,
the interaction of a relativistic wind with an ambient medium and
non-thermal radiation generated at $r\sim r_{\rm dec}$
were considered in many papers (see, for a review \cite{Piran98}).
All these considerations were usually carried out in the frame of
the conventional model which was based on the following assumptions.

(1) Two collisionless shocks form: an external shock that
propagates from the wind front into the ambient medium,
and an internal shock that propagates from the wind front
into the inner wind, with a contact discontinuity at the wind front
between the shocked material.

(2) Electrons are accelerated at the shocks to 
very high energies. 

(3) The shocked matter acquires embedded magnetic fields.
The energy density of these fields is about the energy 
density of high-energy electrons accelerated by the shocks.

(4) Highly accelerated electrons generate high-frequency (X-ray 
and $\gamma$-ray) radiation of GRBs via synchrotron mechanism.

(5)  The efficiency of conversion of the wind energy into 
accelerated electrons and then to high-frequency
radiation of GRBs is about 10\% $\sim$ 30\%. 

The idea about formation of two collisionless shocks 
near the front of a relativistic wind  outflowing from a 
cosmological $\gamma$-ray burster is based mainly on both 
theoretical studies which have shown that 
collisionless shocks can form in a rarified plasma
(e.g., Sagdeev 1962, 1979; \cite{TK71}, \cite{Dawson83}; \cite{Q85})
and the fact that such shocks have been observed 
in the vicinities of a few comets and planets (\cite{Leroy82};
\cite{Livesey}; \cite{Omidi} and references therein). 
Undoubtedly, collisionless shocks exist and can
accelerate electrons to ultrarelativistic energies in
many astrophysical objects such as supernova remnants and
jets of active galactic nuclei. However, for the wind parameters 
which are relevant to cosmological GRBs (see \S~2), formation of 
collisionless shocks in the vicinity of the wind front
is very questionable (e.g., \cite{Smolsky96}; \cite{Mitra}; \cite{Brainerd}).
Indeed, the properties of relativistic, transverse,
magnetosonic collisionless shocks were studied in details by
Gallant et al. (1992) and Hoshino et al. (1992), and it was shown that 
all the upstream ions are electromagnetically reflected from the
shock front. This is similar to reflection of ions from the wind front in our 
simulations. If a shock forms, its thickness is of the order of the gyroradius 
of reflected protons in the upstream magnetic field (e.g., \cite{Hoshino92}). 
In a plasma system, a collisionless shock may form only if
the characteristic size of the system is significantly larger than the shock
thickness. In our case, the radius, $r$, of the wind front is the 
characteristic size of the system. At $r\sim r_{\rm dec}$, an external 
collisionless shock may form in an ambient 
medium just ahead of the wind front if the the gyroradius of protons 
which are reflected from the wind front is significantly smaller than $r_{\rm dec}$. 
In the burster frame, the gyroradius of reflected protons in the ambient 
medium is

\begin{equation}
R_{Bp}={m_pc^2\Gamma_p\over eB_{\rm am}}\simeq 3\times 
10^{16}\left({\Gamma_0\over 10^2}\right)^2\left({B_{\rm am}\over
10^{-6}\,{\rm G}}\right)^{-1}\,\,{\rm cm}\,,
\label{RBp}
\end{equation}
 
\noindent 
where $\Gamma_p\simeq \Gamma_0^2$ is the mean Lorentz factor
of reflected protons and $B_{\rm am}$ is the mean magnetic field of 
the ambient medium. For the interstellar medium in our Galaxy,
the mean magnetic field is about $2\times 10^{-6}$~G (e.g., \cite{MT77}).
If for a GRB the value of $B_{\rm am}$ is $\sim 2\times 10^{-6}$~G, from 
equations (\ref{rdec}) and (\ref{RBp}), for the wind parameters 
which are relevant to cosmological GRBs (see \S~2) we have
$R_{Bp}/r_{\rm dec}\sim 1$ at $\Gamma_0 = 10^2$ and
$R_{Bp}/r_{\rm dec}\sim 10^2$ at $\Gamma_0 = 10^3$.
In this case, an external collisionless shock
cannot form ahead of the wind front, especially if
$\Gamma_0$ is about $10^3$ or even more. Recently, 
Brainerd (1999) came to the same conclusion from another
consideration. As to the internal shock, it cannot form in
a Poynting flux-dominated wind too (e.g., \cite{KFO}; \cite{KC84}).

Our model of the wind -- ambient medium interaction differs 
qualitatively from the conventional model which is based on the assumption
that an external collisionless shock forms just ahead of the wind front.
Although it might seem that observational consequences of our model must 
differ from  observational consequences of the conventional model
significantly, this is not the case. Moreover,
we can see that all the listed assumptions of the conventional model
are confirmed by our simulations, certainly
except of the first assumption on formation of the collisionless shocks.
There is only a modification that LAEMWs are embedded in the region
ahead of the wind front instead of magnetic fields, and 
highly accelerated electrons generate high-frequency emission of GRBs 
via synchro-Compton radiation. However, in our case synchro-Compton 
radiation closely resembles synchrotron radiation. Therefore, 
if in the conventional model of GRB emission from the shocked region 
ahead of the wind front the mechanism of electron
acceleration by a relativistic, collisionless shock is replaced by the
mechanism of electron acceleration at the wind front,
this model will remain otherwise practically unchanged.

At $r >r_{\rm dec}$, the outflowing wind slows down
due to its interaction with the ambient medium, and
when the Lorentz factor of the wind front is about several tens
or less, an external shock may form
just ahead of the front. We believe that the afterglows 
which are observed in $\sim 10^5$~s after some GRBs 
result from acceleration of electrons by such a shock
as it is generally accepted (\cite{Meszaros97b};
\cite{Vietri97}; \cite{Waxman97}; \cite{Wijers97}).

\acknowledgments

This research was supported by MINERVA Foundation, Munich / Germany.

\begin{figure}[p]
  \includegraphics{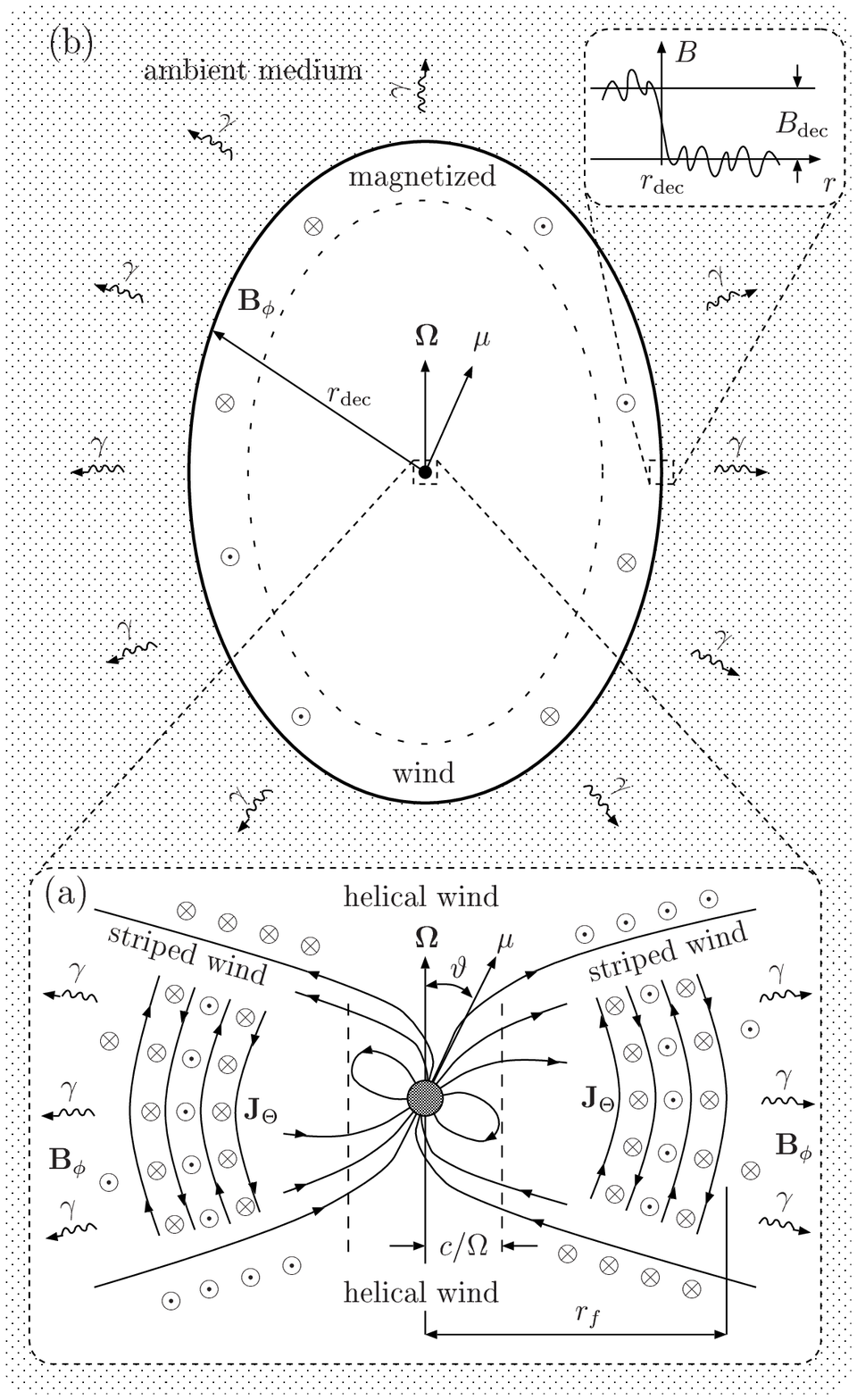}
  \vspace{14cm}
\caption {Sketch (not to scale) of the region where a 
GRB is generated by a relativistic strongly magnetized wind. This wind
is produced by a fast-rotating compact object like a millisecond pulsar with
the angular velocity ${\bf \Omega}$ and the magnetic moment ${\bf \mu}$,
$|\Omega |\sim 10^4$ s$^{-1}$ and 
$|{\bf \mu}|\simeq B_{_S}R^3\sim 10^{34}$ G cm$^3$. The figure part (a)
shows a plausible magnetic topology of the outflowing wind at the
moment when the front of the wind is at the distance $r\sim r_f$ from the
compact object. At this distance, the striped component of the wind field is 
transformed into LAEMWs which decay fast, and $\gamma$-rays are 
generated. At $r\gg r_f$, the magnetic field is helical everywhere in the wind.
The figure part (b) shows the region of the wind --
ambient medium interaction when $r$ is $\sim r_{\rm dec}$.
The main part of the wind energy is concentrated between 
the wind front (thick solid elliptic line) and the dotted elliptic line.
In the right upper corner of (b), it is shown the distribution of magnetic 
fields in the enlarged vicinity of the front.
These fields are a superposition of the wind field which is $B_{\rm dec}
\Theta (r_{\rm dec} -r)$ and the fields of LAEMWs generated at the wind 
front. In the region ahead of the wind front,
electrons of the ambient medium which are accelerated at the front to
extremely high energies move in the fields of LAEMWs and
generate synchro-Compton $\gamma$-rays.
Directions of $\gamma$-ray propagation are shown by arrows with wavy lines.
\label{Fig1}}
\end{figure}

\begin{figure}[p]
  \myfig{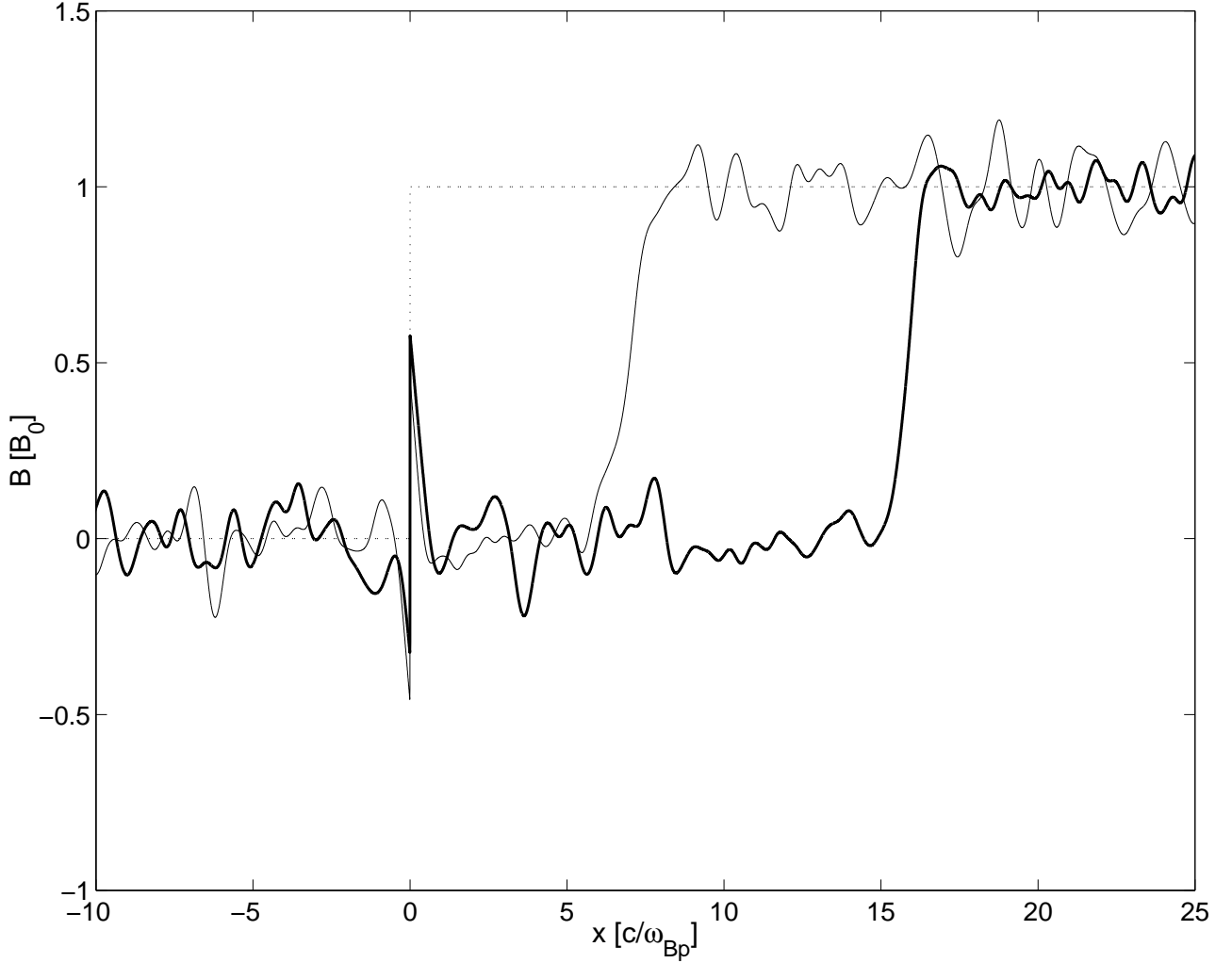}
\caption{Distribution of magnetic field for a simulation
with $B_0=300$ G, $\Gamma_0=300$ and $\alpha =2/3$ at
the moments $t=0$ (dotted line), $t=7.96T_p$ (thin solid line),
and $t=15.9T_p$ (thick solid line). $T_p=2\pi m_pc\Gamma_0/eB_0$ is 
the proton gyroperiod in the magnetic field, $B_0$, of the barrier.
\label{Fig2}}
\end{figure}
 
\begin{figure}[p]
  \myfig{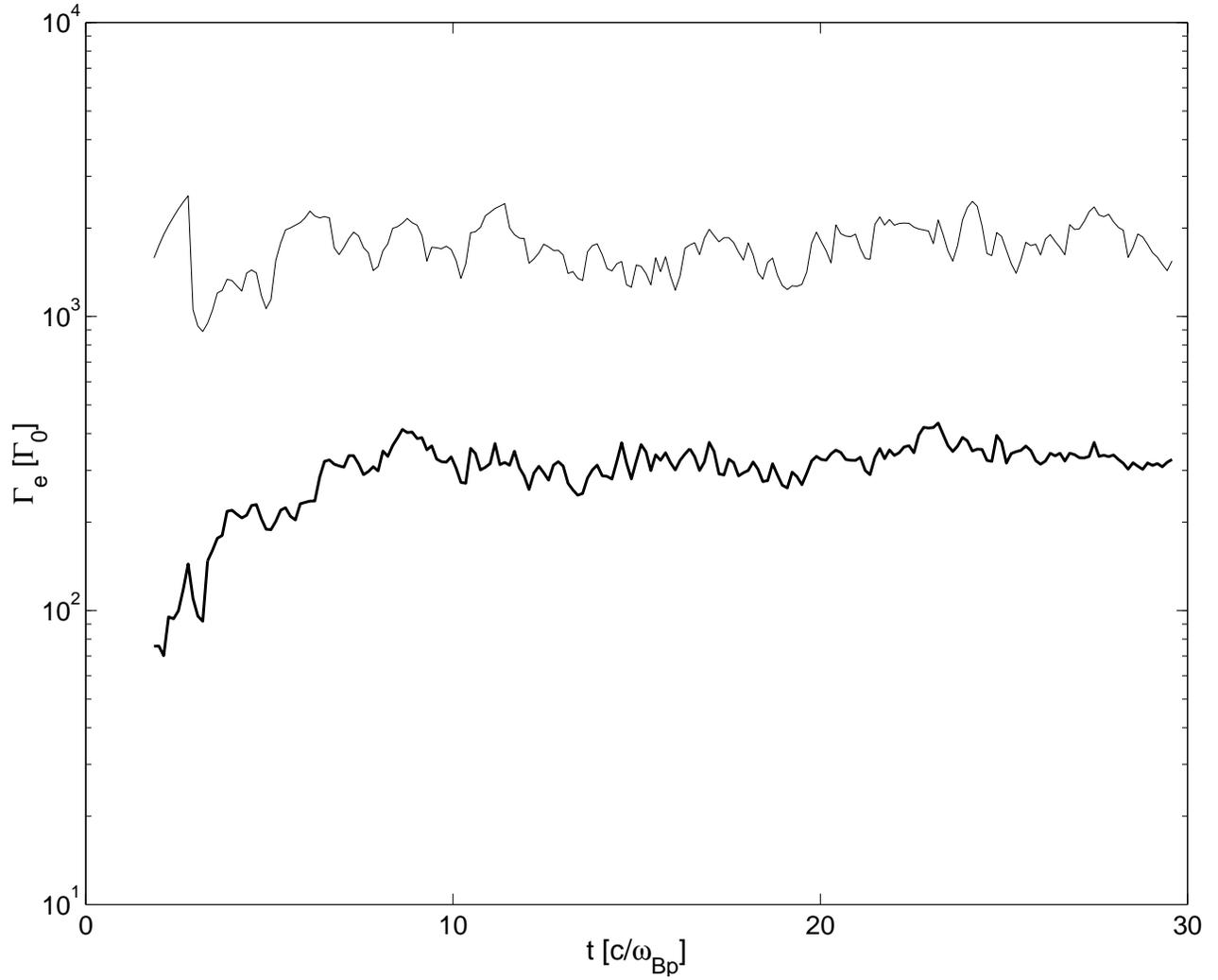}
\caption{Maximum (thin line) and mean (thick line)
energies of outflowing electrons, $v_x<0$, at $x<0$ after thier
reflection from the barrier in a simulation
with $B_0=300$ G, $\Gamma_0=300$ and $\alpha =2/3$.
\label{Fig3}}
\end{figure} 

\begin{figure}[p]
  \myfig{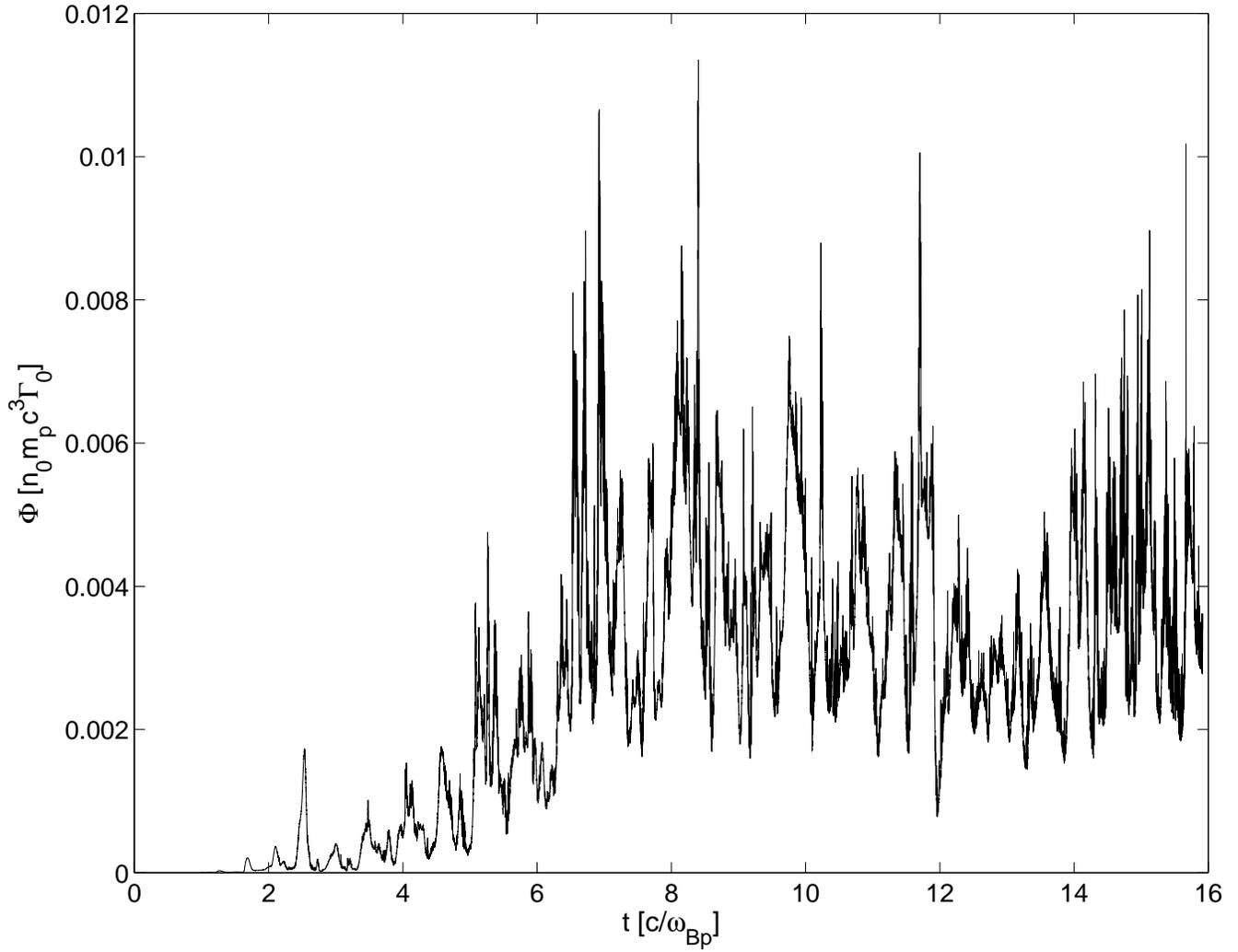}
\caption{Intensity of synchrotron radiation of highly
accelerated electrons per unitary area of the front of the magnetic 
barrier in a simulation
with $B_0=300$ G, $\Gamma_0=300$ and $\alpha =2/3$.
\label{Fig4}}
\end{figure} 

\begin{figure}[p]
  \myfig{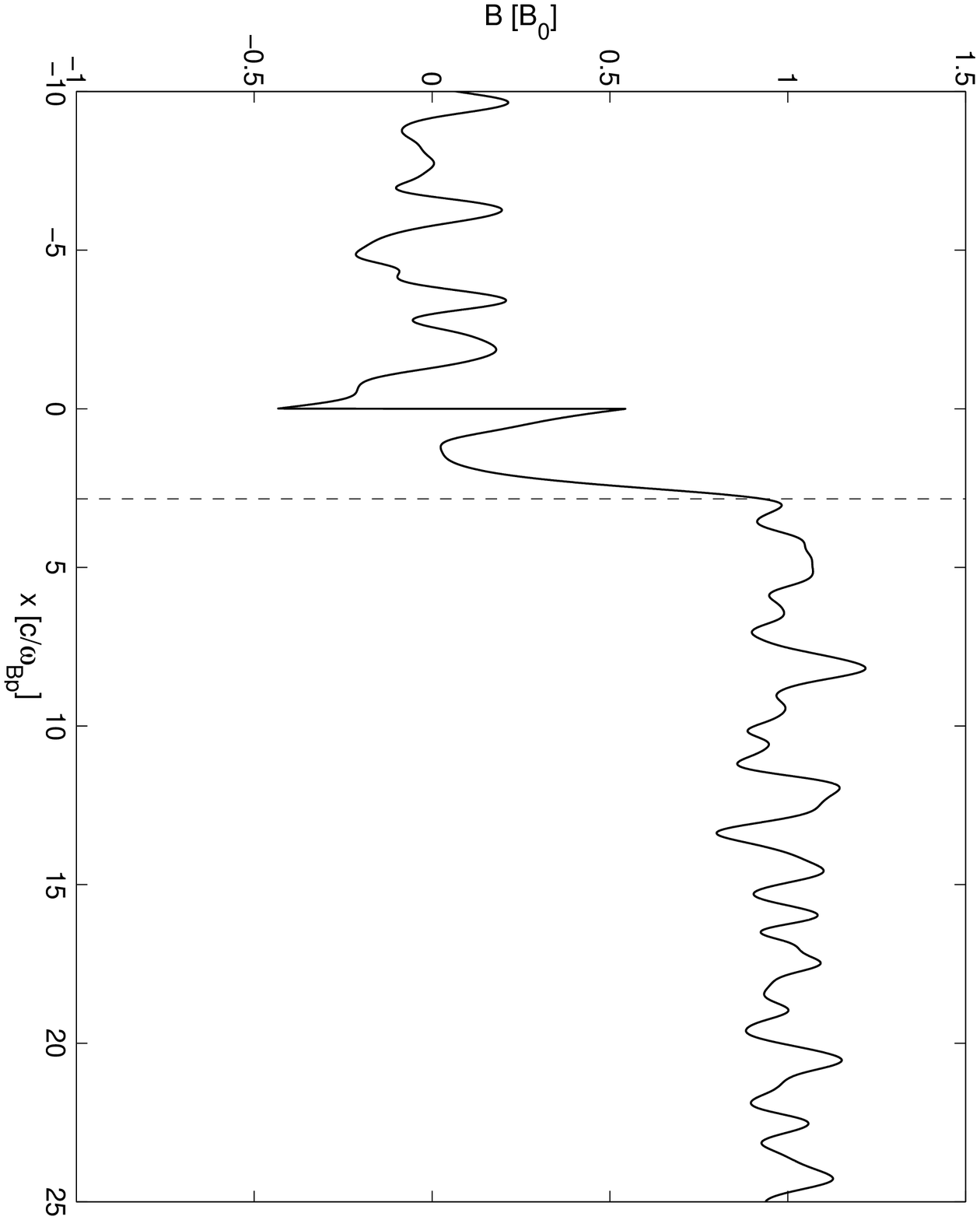}
\caption{Distribution of magnetic field for a simulation
with $B_0=300$ G, $\Gamma_0=300$ and $\alpha =1/2$ at
the moment $t=15.92T_p$. The depth of the beam particle penetration
into the barrier, $x_{\rm pen}=2.84(c/\omega_{Bp})$, is shown by
dashed line.
\label{Fig5}}
\end{figure}
 
\begin{figure}[p]
  \myfig{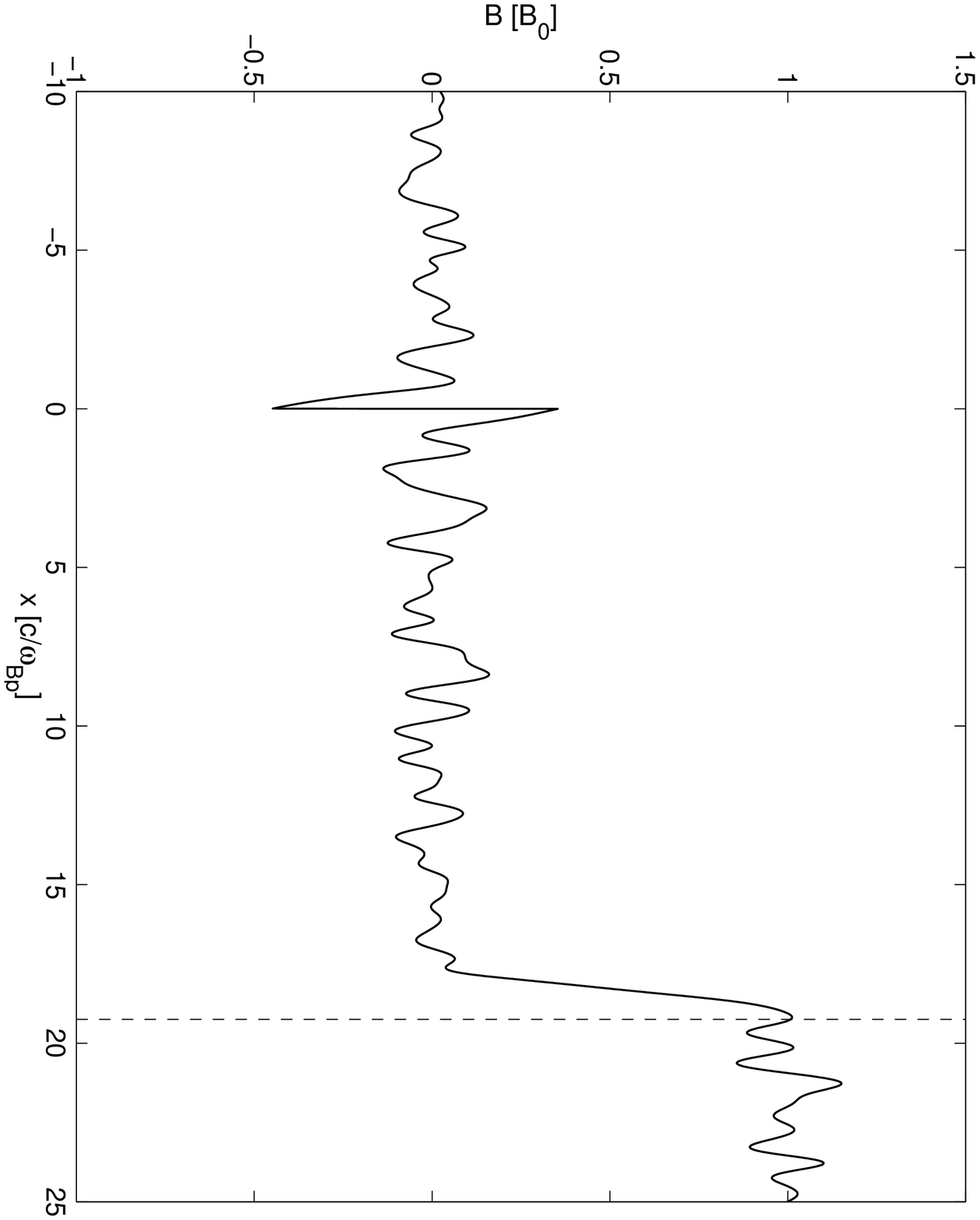}
\caption{Distribution of magnetic field for a simulation
with $B_0=300$ G, $\Gamma_0=300$ and $\alpha =1$ at
the moment $t=15.92T_p$. The depth of the beam particle penetration
into the barrier, $x_{\rm pen}=19.25(c/\omega_{Bp})$, is shown by
dashed line.
\label{Fig6}}
\end{figure}

\begin{figure}[p]
  \myfig{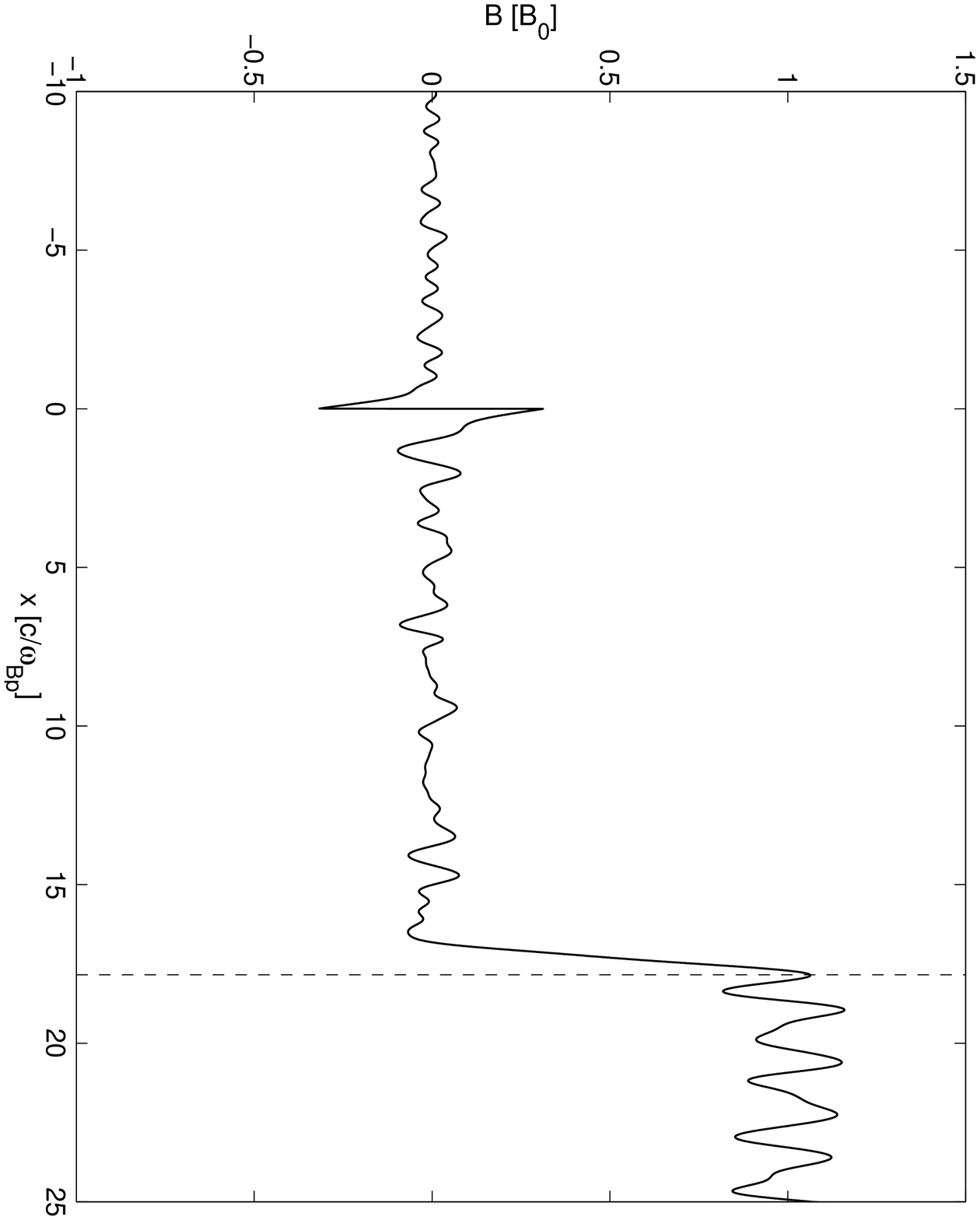}
\caption{Distribution of magnetic field for a simulation
with $B_0=300$ G, $\Gamma_0=300$ and $\alpha =2$ at
the moment $t=7.96T_p$. The depth of the beam particle penetration
into the barrier, $x_{\rm pen}=17.84(c/\omega_{Bp})$, is shown by
dashed line.
\label{Fig7}}
\end{figure}

\begin{figure}[p]
  \myfig{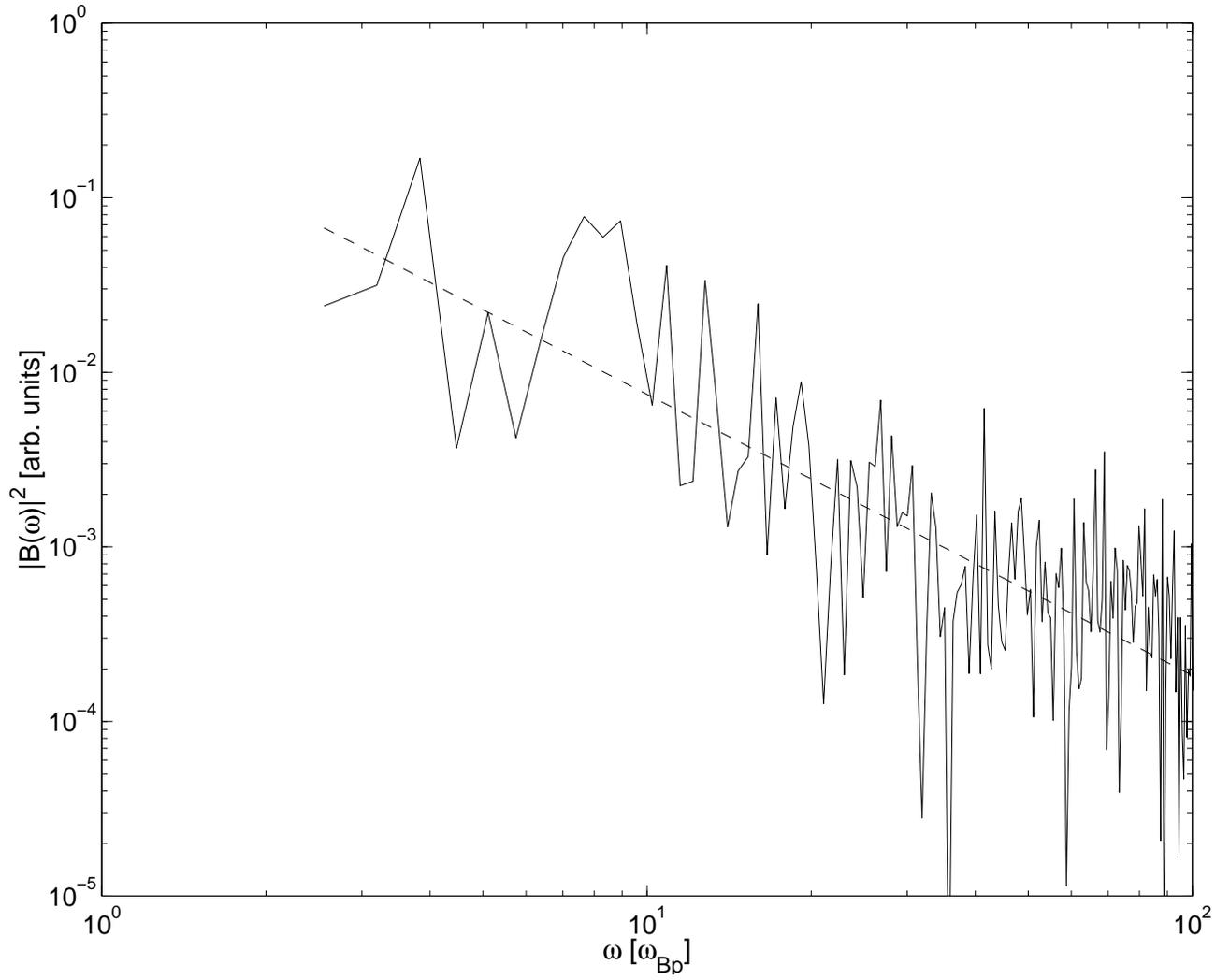}
\caption{Power spectrum of low-frequency electromagnetic waves
generated at the front of the barrier in 
a simulation with $B_0=300$ G, $\Gamma_0=300$ and $\alpha =2/3$.
The spectrum is fitted by a power law (dashed line).
\label{Fig8}}
\end{figure}

\begin{figure}[p]
  \myfig{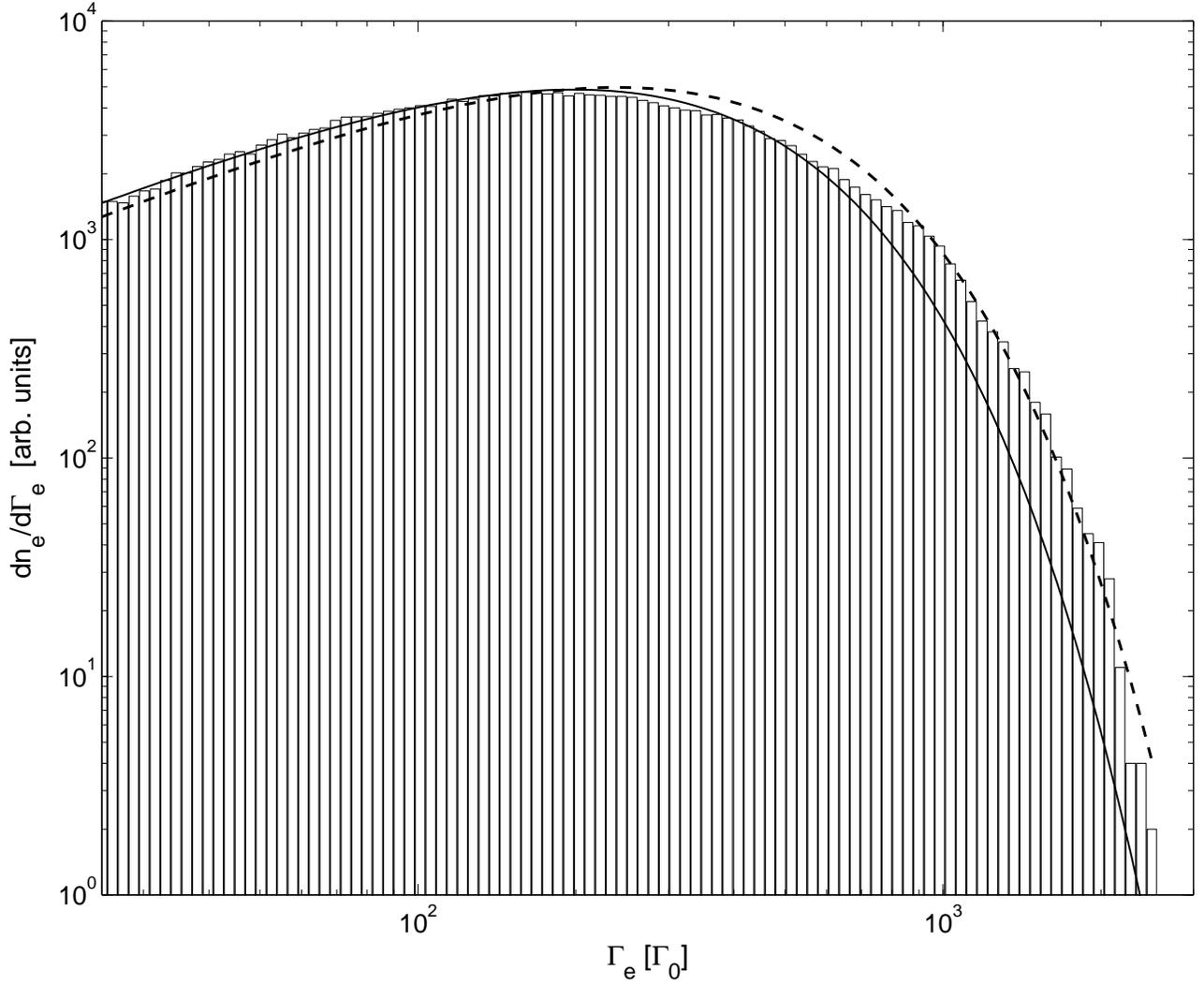}
\caption{Energy spectrum of highly accelerated electrons
in the region ahead of the wind front in the frame of the wind
for a simulation with $B_0=300$ G, $\Gamma_0=300$ and $\alpha =2/3$.
The electron spectrum is fitted by a two-dimensional relativistic
Maxwellian with a relativistic temperature $T=m_ec^2\Gamma_{_T}/k$,
where $\Gamma_{_T}$ is equal to either $240\Gamma_0$
(dashed line) or $200\Gamma_0$ (solid line).
\label{Fig9}}
\end{figure}

\begin{figure}[p]
  \myfig{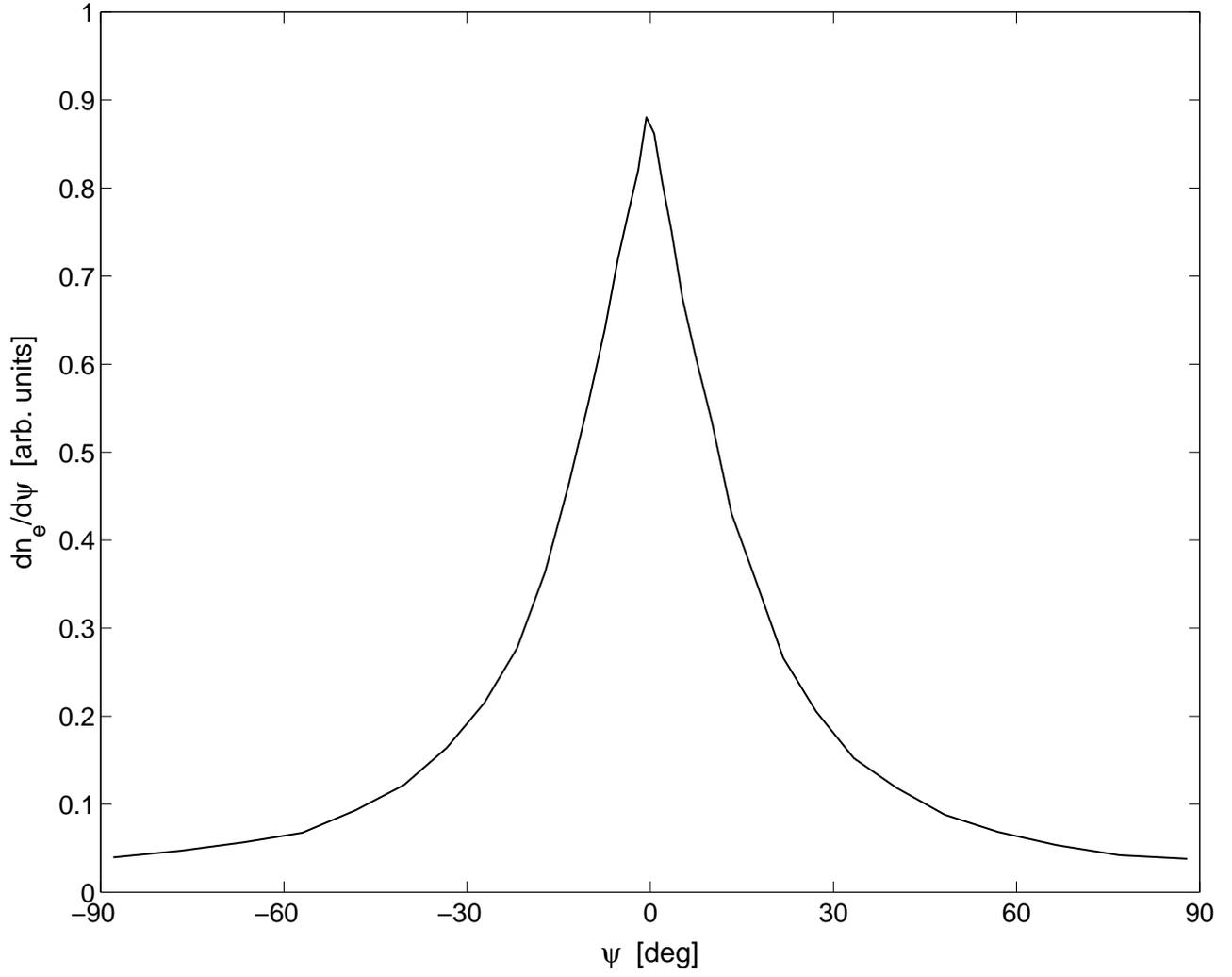}
\caption{Angular distribution of outflowing high-energy electrons 
in a simulation with $B_0=300$ G, $\Gamma_0=300$ and $\alpha =2/3$.
\label{Fig10}}
\end{figure}

\begin{figure}[p]
  \myfig{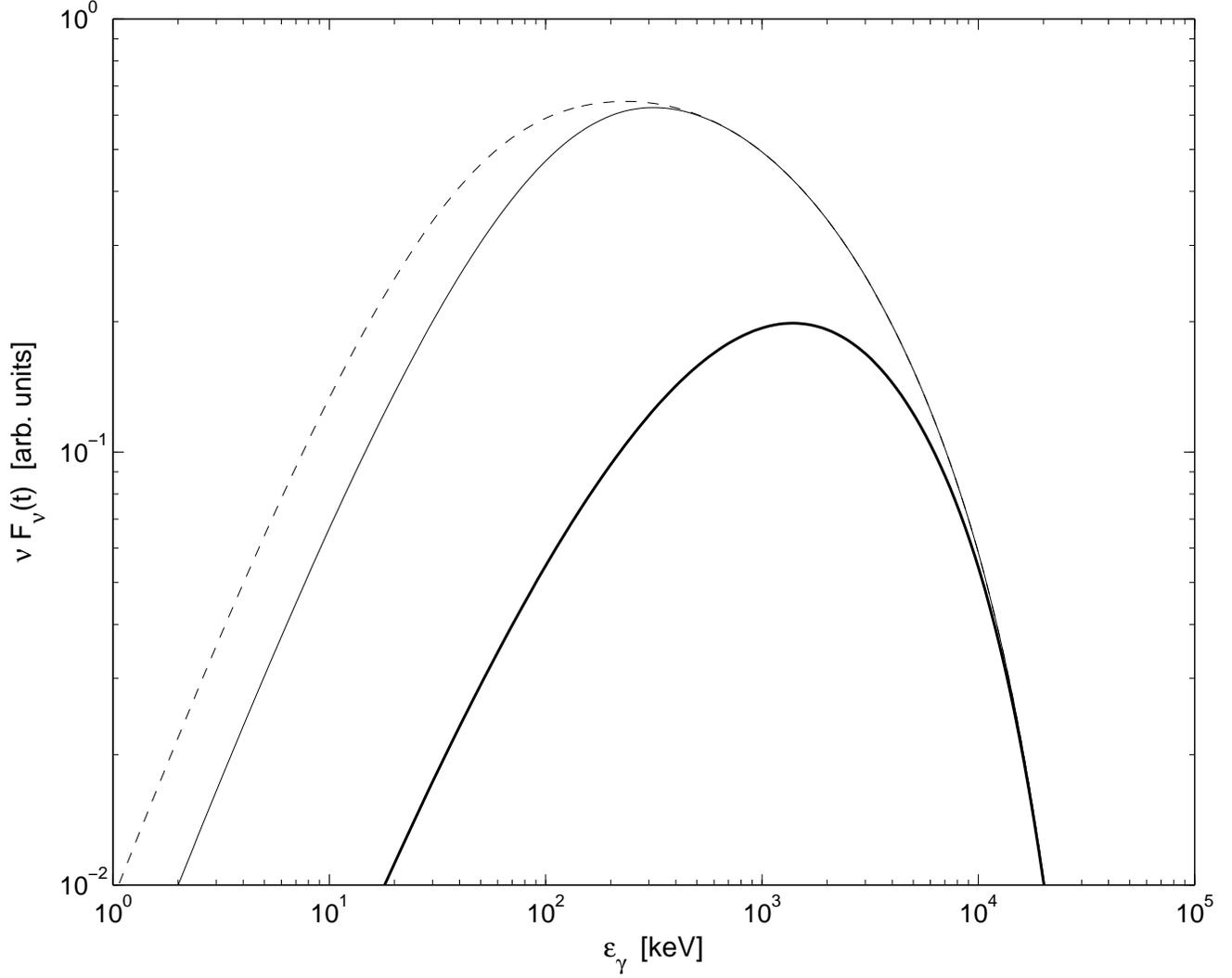}
\caption{Calculated spectral power of synchro-Compton radiation 
from the region ahead of the wind front as a function of photon energy
for $\Gamma_0 = 150$, $B_0= 300$ G, $\langle B_w\rangle =
0.1 B_0$, $\Gamma_T = 200\Gamma_0 = 3\times 10^4$ and $z=1$.
The spectrum of radiation is given for the moments when the fraction of
the energy of high-energy electrons injected into the region ahead of 
the wind front which is radiated is nearly zero (thick solid line), 36\% (thin solid line) and
58\% (dashed line).
\label{Fig11}}
\end{figure}

\end{document}